\documentclass[12pt]{article}
\usepackage{hyperref}
\usepackage[
    type={CC},
    modifier={by},
    version={4.0},
]{doclicense}
\usepackage[utf8]{inputenc}
\usepackage[a4paper, total={6.5in, 9.5in}]{geometry}
\usepackage{enumitem}
\usepackage[reqno]{amsmath}
\counterwithin{equation}{section}
\usepackage{exscale} 
\usepackage{fontenc}
\usepackage{makeidx}
\usepackage{amsfonts}
\usepackage{amstext}
\usepackage{amsrefs}
\usepackage{amsthm}
\usepackage{amssymb}
\usepackage{bbm}
\usepackage[utf8]{inputenc}
\usepackage{xcolor}
\usepackage{float}
\usepackage{tikz} 
\usepackage{appendix}
\usepackage{booktabs}
\usepackage{float}
\usepackage{esint}
\usepackage{authblk}
\usepackage{fancyhdr}
\usepackage{mathrsfs}
\usepackage{array}
\usepackage{multirow}
\usepackage{authblk}

\newcommand{\R}{\mathbb{R}}

\newcommand{\D}{\Delta}

\newcommand{\jnabla}{{\langle\nabla\rangle}}
\newcommand{\dXY}{{\mathrm{dist}(X,Y)}}
\newcommand{\Teps}{{G_\varepsilon}}
\newcommand{\Tzero}{{G_0}}

\theoremstyle{plain}
\newtheorem{theo}{Theorem}[section]
\newtheorem{prop}[theo]{Proposition}
\newtheorem{lemma}[theo]{Lemma}
\newtheorem{cor}[theo]{Corollary}

\newtheorem{rem}[theo]{Remark}

\newtheorem{definition}[theo]{Definition}

\newcommand{\diff}{\mathrm{d}}

\allowdisplaybreaks

\title{Exponential Decay outside of the Light Cone for the Pseudo-Relativistic Non-Autonomous Schrödinger Equation}
\date{}
\author{S\'ebastien Breteaux}
\author{J\'er\'emy Faupin}
\author{Viviana Grasselli}
\affil{Université de Lorraine, CNRS, IECL, F-57000 Metz, France}

\begin{document}

\maketitle

\begin{center}
\textit{Dedicated to Israel Michael Sigal on the occasion of his 80th birthday}
\end{center}

\begin{abstract}
    We establish a maximal velocity bound for a pseudo-relativistic quantum particle in an external time-dependent potential. Our estimate shows that the probability for the particle, starting in a convex set $X\subset\mathbb{R}^d$ at $t=0$, to reach a convex set $Y\subset\mathbb{R}^d$ at a time $t>0$, is bounded by $e^{-2\delta}$ where $\delta$ is the distance from $Y$ to the section at time $t$ of the light cone generated by $X$.
\end{abstract}

\section{Introduction}
We consider a pseudo-relativistic quantum particle placed in a time-dependent external potential $V_t$, in dimension $d\ge1$. In units such that the Planck constant and the velocity of light are equal to $1$, the dynamics of the particle is given by the linear pseudo-relativistic (or semi-relativistic) Schrödinger equation
\begin{equation*}
\begin{cases}
    i\partial_t\psi_t=\big(\sqrt{-\Delta+m^2}-m+V_t\big)\psi_t\\
    \psi_{t=0}=\psi_0,
\end{cases}
\end{equation*}
in $\mathbb{R}^d$, where $m>0$ is the mass of the particle. For all $t\ge0$, $V_t:\mathbb{R}^d\to\mathbb{R}$ is a time-dependent real-valued potential representing the external forces applied to the particle, and the kinetic energy $\sqrt{-\Delta+m^2}-m$ is the Fourier multiplier defined by
\begin{equation*}
    \sqrt{-\Delta+m^2}-m=\mathcal{F}^{-1}\big(\sqrt{|\xi|^2+m^2}-m\big)\mathcal{F}.
\end{equation*}
Here and it what follows, $\mathcal{F}$ stands for the usual Fourier transform (we omit the choice of the normalization of the Fourier transform as it is irrelevant for our purpose). To simplify the exposition, we drop the constant mass term, which will have no influence on our results, and we take $m=1$. The main equation studied in this paper is therefore
\begin{equation}\label{Cauchy-pb-propagateur-Schroedinger}
\begin{cases}
    i\partial_t\psi_t=(\langle\nabla\rangle+V_t)\psi_t, \qquad \jnabla := \sqrt{1-\Delta}, \\
    \psi_{t=0}=\psi_0.
\end{cases}
\end{equation}
We will assume that the initial state $\psi_0$ belongs to the Sobolev space $H^{1/2}(\mathbb{R}^d)$.

\begin{figure}[H]\label{fig:cone-lumiere}
    \begin{center}
         \includegraphics[angle=0,width=15cm,origin=c]{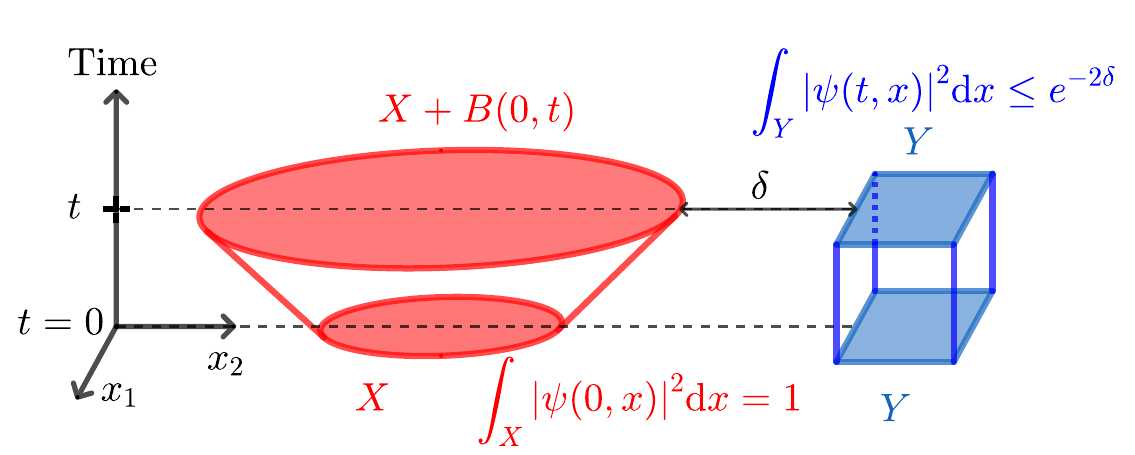}
    \caption{Consider $X$ and $Y$ convex subsets of $\mathbb{R}^d$ ($d=2$ in the illustration). Let $B(0,t)$ be the closed ball of radius $t>0$ centered at the origin. If the section $X+B(0,t)$ of the light cone generated by $X$ at a time $t$ is at a distance $\delta$ from $Y$, then a particle in $X$ at time $0$ will reach $Y$ at time $t>0$ with a probability lower than $e^{-2\delta}$.}
    \end{center}
\end{figure}

Our main concern is the estimate of the speed of propagation of solutions to \eqref{Cauchy-pb-propagateur-Schroedinger}. More precisely, we aim at establishing a maximal velocity bound, showing that the probability that a quantum particle whose dynamics is described by \eqref{Cauchy-pb-propagateur-Schroedinger} travels faster than the speed of light is exponentially small as illustrated intuitively in Fig.~\ref{fig:cone-lumiere}. This will be made more precise in the statements below.

Many works have been devoted to estimating the speed of propagation for quantum systems in various contexts in the last years. In the eighties and the nineties, propagation estimates for non-relativistic $N$-body quantum systems became a central tool in the study of their scattering theory, in particular for the proof of Asymptotic Completeness of the wave operators \cite{Enss83,SigalSoffer88,SigalSoffer90,Graf90,Gerard92,Skibsted91,Derezinski93,HunzikerSigalSoffer99}, see also \cite{derezinski_ger} for a textbook presentation. Propagation estimates were next extended to models related to non-relativistic quantum electrodynamics, again in relation with their scattering theory, e.g. in~\cite{DerezinskiGerard99,FrohlichGriesemerSchlein02,BonyFaupinSigal12,FaupinSigal14}. More recently, the maximal velocity of general non-relativistic two-body quantum systems was estimated in~\cite{ArbunichPusateriSigalSoffer21}, using a method based on differential inequalities sometimes  called the ASTLO method (Adiabatic Spacetime Localization Observables). The latter was then refined and successfully applied in various settings: see~\cite{FaupinLemmSigal22_LR,FaupinLemmSigal22_maxv,LemmRubilianiZhang23,LemmRubilianiZhang25,KuwaharaLemmRubiliani25} for Bose-Hubbard type Hamiltonians in relation with the celebrated Lieb-Robinson bounds~\cite{LiebRobinson}; see~\cite{BreteauxFaupinLemmSigal22,BreteauxFaupinLemmSigalYangZhang24} for the Lindblad master equation describing the effective evolution of open quantum systems; see also~\cite{ArbunichFaupinPusateriSigal23} for the non-linear Hartree equation describing the effective dynamics of many-body quantum systems. In~\cite{SigalWU}, a novel method to prove maximal velocity bounds for quantum systems was introduced, combining a clever conjugation of the quantum evolution by a family of unitary operators and an analyticity argument. The method was then applied to open quantum systems in~\cite{SigalWu25} and to quantum lattice gases in~\cite{FaupinLemmSigalZhang25}. Our approach in this paper is strongly inspired by~\cite{SigalWU}, with some differences  that will be detailed below.

Now we describe our main result in precise terms. As usual, for non-autonomous Schrödin\-ger-type equations, given a family of real-valued potentials $(V_t)_{t\in[0,T]}$, the existence of solutions to \eqref{Cauchy-pb-propagateur-Schroedinger} on the time interval $[0,T]$ is ensured by the existence of a unitary dynamics~$(U(t,s))_{t,s\in[0,T]}$ generated by $(\jnabla+V_t)_{t\in[0,T]}$. Depending on the context, different definitions of unitary dynamics are considered in the literature. In this paper, a unitary dynamics generated by $(\jnabla+V_t)_{t\in[0,T]}$ is defined as a family of unitary operators $(U(t,s))_{t,s\in[0,T]}$ on~$L^2(\mathbb{R}^d)$ such that, for all $t,s,r$ in~$[0,T]$, $U(t,t)=\mathbf{1}_{L^2}$, $U(t,s)U(s,r)=U(t,r)$ and
\begin{equation}
    \forall\psi\in H^{\frac12}(\mathbb{R}^d), \qquad i\partial_tU(t,s)\psi=(\jnabla+V_t)\,U(t,s)\,\psi,
\end{equation}
where the equality in the previous equation holds in $H^{-1/2}(\mathbb{R}^d)$. We refer to the next section for a more precise and more general definition of unitary dynamics in our setting. To simplify the notations, we will write $U_t=U(t,0)$ and say that $(U_t)_{t\in[0,T]}$ is the propagator generated by $(\jnabla+V_t)_{t\in[0,T]}$.

Let $\|\cdot\|_{L^2}$ be the operator norm on $L^2(\mathbb{R}^d)$ and let $\mathrm{dist}(X,Y)$ be the distance between two subsets $X$ and $Y$ of $\mathbb{R}^d$.  Our main result provides, for any $t$ in $[0,T]$, an estimate on the norm $\|\mathbf{1}_YU_t\mathbf{1}_X\|_{\mathcal{B}(L^2)}$ of the unitary propagator generated by $(\jnabla+V_t)_{t\in[0,T]}$ composed with the characteristic functions $\mathbf{1}_X$ and $\mathbf{1}_Y$ of any convex subsets~$X$ and $Y$ of $\mathbb{R}^d$:
\begin{theo}\label{th:max-vel}
Let $T>0$ and $(V_t)_{t\in[0,T]}$ be a family of real-valued potentials such that the family~$(U_t)_{t\in[0,T]}$ is a propagator generated by $(\langle\nabla\rangle+V_t)_{t\in[0,T]}$. If $X$ and $Y$ are convex subsets of~$\mathbb{R}^d$, then
\begin{equation}\label{eq:main-est}
        \forall t\in [0,T]\,,\quad \|\mathbf{1}_Y U_t \mathbf{1}_X\|_{\mathcal{B}(L^2)} \le e^{t-\dXY}.
    \end{equation}
\end{theo}

\begin{rem}
We recall that the speed of light and the mass of the particle are equal to $1$ in our units. For the relativistic dispersion relation $\sqrt{-c^2\Delta+m^2c^4}-mc^2$, instead of \eqref{eq:main-est}, our proof gives
    \begin{equation*}
        \|\mathbf{1}_YU_t\mathbf{1}_X\|_{\mathcal{B}(L^2)}\le e^{mc^2(ct-\dXY)}.
    \end{equation*}
\end{rem}
As mentioned before, Theorem \ref{th:max-vel} should be interpreted as a maximal velocity estimate for the propagation of a semi-relativistic quantum particle in the time-dependent external potential $V_t$. Indeed, if one considers a unit vector~$\psi_0$ in $L^2(\mathbb{R}^d)$ satisfying $\psi_0 = \mathbf{1}_X \psi_0$, then Theorem \ref{th:max-vel} yields 
\[
\|\mathbf{1}_Y \psi_t \|_{L^2} = \|\mathbf{1}_Y U_t \mathbf{1}_X \psi_0 \|_{L^2} \leq \|\mathbf{1}_Y U_t \mathbf{1}_X\|_{\mathcal{B}(L^2)}\|\mathbf{1}_X \psi_0\|_{L^2}\leq e^{t-\dXY}\,.
\]
In other words, if the position of the particle is initially, at time $t=0$, localized in $X$, then the probability that the particle is in $Y$ at time $t>0$ is smaller than $e^{2(t-\dXY)}$. Therefore the probability that the particle travels faster than the speed of light (equal to $1$ in our units) between any convex subsets $X$ and $Y$ of $\mathbb{R}^d$ is exponentially small. A key feature of our estimate, compared to previous results, is that it shows that the probability for the particle, starting in $X$, to be in a convex region $Y$ outside of the section of the light cone at time $t>0$ (namely $Y\subset\mathbb{R}^d\setminus (X+B(0,t))$), is exponentially small, more precisely bounded by $e^{-2\delta}$ where $\delta>0$ is the distance from $Y$ to $X+B(0,t)$. In previous works (see in particular \cite{SigalWU}), the obtained maximal velocity estimates are typically of the form $\|\mathbf{1}_Y U_t \mathbf{1}_X\|_{\mathcal{B}(L^2)}\leq C_{\mu,c} e^{\mu(ct- \dXY)}$ for some $C_{\mu,c}>1$ and any $\mu<1$ and $c>1$, without explicit control on $C_{\mu,c}$. To our knowledge, Theorem \ref{th:max-vel} is the first result providing a maximal velocity estimate for a quantum particle in the continuum with such a uniform control in the distance to the light cone. However, the result in \cite{SigalWU} holds for more general kinetic energy $\omega(-i\nabla)$, with suitable assumptions on $\omega$.

We have the following remark concerning the sharpness of our result:

\begin{rem}\label{rk:sharp}
 The exponentially small error term in the maximal velocity estimate \eqref{eq:main-est} is ``sharp'' for $t\le\mathrm{dist}(X,Y)$ in the sense that:
\begin{enumerate}
    \item If $C<1$, there exist convex subsets $X$ and $Y$ such that the estimate 
\begin{equation*}
        \forall t\in [0,T]\,,\quad \|\mathbf{1}_Y U_t \mathbf{1}_X\|_{\mathcal{B}(L^2)} \le C e^{t-\dXY},
\end{equation*}
does \emph{not} hold. This is obvious since, if $X\cap Y$ is a subset with a positive Lebesgue measure, then $\|\mathbf{1}_Y U_t \mathbf{1}_X\|_{\mathcal{B}(L^2)}=1$ at $t=0$.
\item If $c<1$, there exist convex subsets $X$ and $Y$ and a time-dependent potential $V_t$ such that the estimate 
\begin{equation*}
        \forall t\in [0,T]\,,\quad \|\mathbf{1}_Y U_t \mathbf{1}_X\|_{\mathcal{B}(L^2)} \le e^{ct-\dXY},
\end{equation*}
does \emph{not} hold. This statement is proven in the case of the free evolution, $V_t=0$, in Appendix \ref{app:sharp} (see Corollary \ref{cor:sharp}).
\end{enumerate}
\end{rem}

As in \cite{SigalWU} (see also \cite{FaupinLemmSigalZhang25}), the idea of the proof of Theorem~\ref{th:max-vel} is to construct a suitable function $\ell$ such that 
\begin{align}
\|\mathbf{1}_Y U_t \mathbf{1}_X\|_{\mathcal{B}(L^2)}
    \le \underset{\le\, \exp\big(-\frac{\dXY}{2}\big)}{\underbrace{\|\mathbf{1}_Ye^{\ell(x)}\|_{\mathcal{B}(L^2)}}} \,
    \underset{\le\, \exp(t)}{\underbrace{\|e^{-\ell(x)}U_te^{\ell(x)}\|_{\mathcal{B}(L^2)}}}\,
    \underset{\le\, \exp\big(-\frac{\dXY}{2}\big)}{\underbrace{\|e^{-\ell(x)}\mathbf{1}_X\|_{\mathcal{B}(L^2)}}}
   \,.\label{eq:idee-borne-de-vitesse-maximale}
\end{align}
Our choice of the function $\ell$ is very similar to that of \cite{SigalWU,FaupinLemmSigalZhang25}. However, instead of using an analyticity argument as in \cite{SigalWU}, we introduce a bounded approximation $\ell_\varepsilon$ of $\ell$, for small~$\varepsilon>0$, establish various mapping properties of the transformed kinetic energy operator~$e^{\ell_\varepsilon(x)}\jnabla e^{-\ell_\varepsilon(x)}$, and then take the limit $\varepsilon\to0$. A careful analysis, using the explicit form of the pseudo-relativistic kinetic energy, then allows us to reach the ``sharp'' estimate \eqref{eq:main-est}.

Theorem \ref{th:max-vel} can be extended to non necessarily convex subsets $X$ and $Y$ at the price of losing the sharpness of the exponential decay: an additional multiplicative constant and a polynomial term in the distance between $X$ and $Y$ appear in the error term.
\begin{cor}\label{cor:max-vel-bounds-general-subsets}There exists $C_d>0$ such that, if   $(V_t)_{t\in[0,T]}$, with $T>0$, is a family of real-valued potentials such that $(U_t)_{t\in[0,T]}$ is a propagator generated by $(\langle\nabla\rangle+V_t)_{t\in[0,T]}$, then, for any Borel subsets  $X$ and $Y$ of $\mathbb{R}^{d}$ the bound
\begin{equation}\label{eq:main-est-gen}
\forall t\in[0,T]\,,\qquad \|\mathbf{1}_{Y}U_{t}\mathbf{1}_{X}\|_{\mathcal{B}(L^{2})}\leq C_{d} \, e^{t-\dXY} \,  \langle\dXY\rangle^{d}\,,
\end{equation}
holds, with $\langle r \rangle = \sqrt{1+r^2}$.
\end{cor}

The error term in \eqref{eq:main-est-gen} may be improved, at least if one considers specific subsets $X$ and~$Y$, e.g. if $X$ is a ball and $Y$ the complement of a larger ball, both centered at the origin. We do not elaborate here.

\begin{rem}In a companion paper \cite{BFG2b}, we estimate the speed of propagation for the non-linear pseudo-relativistic Hartree equation (the boson star equation). For non-convex subsets~$X$ and~$Y$, we show that we can apply Corollary \ref{cor:max-vel-bounds-general-subsets} (and Proposition~\ref{cor:existence-unitary-propagator-for-time-dependent-potential} below) with the ``state-dependent'' potential $V_t=w*|\psi_t|^2$ under suitable conditions on $w$ and the initial state~$\psi_0$ (here $w$ is a suitable convolution potential and~$\psi_t$ is the solution to the pseudo-relativistic Hartree equation).
\end{rem}

Our main results, Theorem \ref{th:max-vel} and Corollary \ref{cor:max-vel-bounds-general-subsets}, are stated under the assumption that~$(\jnabla+V_t)_{0\le t\le T}$ generates a unitary propagator. The following proposition provides a simple criterion ensuring that this assumption is satisfied. 

The space of bounded operators from the Sobolev spaces $H^{1/2}(\mathbb{R}^d)$ to $H^{-1/2}(\mathbb{R}^d)$ is denoted by $\mathcal{B}(H^{1/2},H^{-1/2})$ and endowed with the norm
\begin{equation*}
    \|B\|_{\mathcal{B}(H^{\frac12},H^{-\frac12})}:=\big\|\jnabla^{-\frac12}B\jnabla^{-\frac12}\big\|_{\mathcal{B}(L^2)},
\end{equation*}
where ${\mathcal{B}(L^2)}$ stands for the set of bounded operators on $L^2(\mathbb{R}^d)$. 
As usual, we will identify a function $V:\mathbb{R}^d\to\mathbb{R}$ and the multiplication operator associated to it. 
We recall that the precise notion of unitary propagators that we consider in this paper will be given in Section~\ref{sec:Notations-Definitions}.
\begin{prop}\label{cor:existence-unitary-propagator-for-time-dependent-potential}
Let $T>0$.
Suppose that, for all $t$ in $[0,T]$, $V_t:\mathbb{R}^d\to\mathbb{R}$ decomposes as~$V_t=V_{\infty,t}+V_{\mathcal{B},t}$ with $V_{\infty,t}$ in $L^\infty$,~$V_{\mathcal{B},t}$ in $\mathcal{B}(H^{1/2},H^{-1/2})$ and
\begin{enumerate}
    \item \label{item:borne-Vd} for all $t$ in $[0,T]$,  $\|V_{\mathcal{B},t}\|_{\mathcal{B}(H^{1/2},H^{-1/2})} <1$\,, 
    \item \label{item:borne-Vinfty} $\sup_{t\in [0,T]}\|V_{\infty,t}\|_{L^\infty}<\infty$\,,
    \item \label{item:borne-dtV}$ \sup_{t\in [0,T]}\|\partial_t V_t\|_{\mathcal{B}(H^{1/2},H^{-1/2})+L^\infty} < \infty$\,.
\end{enumerate}
Then $(\jnabla+V_t)_{t\in[0,T]}$ generates a unitary propagator.
\end{prop}

\begin{rem}
The ``smallness'' condition $\|V_{\mathcal{B},t}\|_{\mathcal{B}(H^{1/2},H^{-1/2})} <1$ ensures that for all~$t$ in~$[0,T]$, the operator $\jnabla + V_t$ identifies to a self-adjoint operator with form domain $H^{1/2}$, by the KLMN Theorem. See Section \ref{sec:Unitary} below.
\end{rem}

\begin{rem}
    For $d\ge2$, we have $L^{d,\infty}\subset \mathcal{B}(H^{1/2},H^{-1/2})$, with $L^{d,\infty}$ the usual weak Lebesgue space. (Recall that  a function in $L^{d,\infty}$ is identified with the associated multiplication operator.) Indeed, if $V$ belongs to $L^{d,\infty}$, then for all $\psi\in\dot{H}^{1/2}$,
    \begin{equation*}
    \Big|\int_{\mathbb{R}^d}V|\psi|^2\Big| 
    \lesssim \|V\|_{L^{d,\infty}} \|\psi\|_{L^{\frac{2d}{d-1},2}}^2
   \lesssim \|V\|_{L^{d,\infty}} \|\psi\|_{\dot{H}^{1/2}}^2,
   \end{equation*}
   by Hölder's inequality and the Sobolev embedding in Lorentz spaces (see e.g. \cite{LemarieRieusset}*{Chapter 2} for the definition and some properties of Lorentz spaces). The conclusion of Proposition \ref{cor:existence-unitary-propagator-for-time-dependent-potential} therefore also holds if we assume that, for all $t$ in $[0,T]$, $V_t=V_{d,t}+V_{\infty,t}$ with $V_{d,t}$ in $L^{d,\infty}$, $V_\infty$ in $L^\infty$ and
   \begin{enumerate}
    \item \label{item:borne-Vd-rk} for all $t$ in $[0,T]$, $\|V_{d,t}\|_{L^{d,\infty}} <1/K_d$\,, 
    \item \label{item:borne-Vinfty-rk} $\sup_{t\in [0,T]}\|V_{\infty,t}\|_{L^\infty}<\infty$\,,
    \item \label{item:borne-dtV-rk}$ \sup_{t\in [0,T]}\|\partial_t V_t\|_{L^{d,\infty}+L^\infty} < \infty$\,,
\end{enumerate}
with
\begin{equation*}
K_d:=\sup_{\substack{V\in L^{d,\infty}\setminus\{0\}\\\psi \in \dot{H}^{1/2}\setminus\{0\}}} \frac{ \big|\int_{\mathbb{R}^d}V|\psi|^2\big|}{\|V\|_{L^{d,\infty}} \|\psi\|_{\dot{H}^{1/2}}^2}\,.
\end{equation*}
\end{rem}

\begin{rem}
To prove Proposition \ref{cor:existence-unitary-propagator-for-time-dependent-potential}, we use an abstract criterion proven in \cite{AmmariBreteaux} to ensure the existence of a unitary propagator generated by a family of time-dependent self-adjoint operators in a Hilbert space. We emphasize that other criteria have been used in the literature, see e.g. \cite{derezinski_ger}*{Appendix B.3} or \cite[Chapter 5]{Pazy}. Applying these different criteria would give different classes of admissible potentials for our results to hold, which would neither contain nor be contained in the class of potentials we obtain in Proposition \ref{cor:existence-unitary-propagator-for-time-dependent-potential}. Advantages of using the criterion of \cite{AmmariBreteaux} are, first, that we can cover the usual class of time-independent potentials and second, that the conditions imposed on the potential $V_t$ do not involve space derivatives of $V_t$.
\end{rem}

\paragraph{Organisation of the Paper.}
In Section~\ref{sec:Notations-Definitions} we introduce the notations we will be using, and we state in precise terms the notion of unitary propagator that we consider. Section~\ref{sec:max-vel-estimates} presents the proofs of Theorem \ref{th:max-vel} and Corollary \ref{cor:max-vel-bounds-general-subsets} on maximal velocity estimates. Our proofs are essentially self-contained, they only rely on standard results which can be found for instance in \cite{ReedSimon1,reed_sim_vol2}. Section~\ref{sec:Unitary} is devoted to the existence and uniqueness of unitary propagators, in particular the proof of Proposition \ref{cor:existence-unitary-propagator-for-time-dependent-potential}. It relies on some previous abstract results from~\cite{AmmariBreteaux}. 
In Appendix \ref{app:sharp}, we justify the ``sharpness'' of the maximal velocity estimate~\eqref{eq:main-est} as mentioned in Remark \ref{rk:sharp}.

\section*{Acknowledgements}
This research was funded, in whole or in part, by l’Agence Nationale de la Recherche
(ANR), project ANR-22-CE92-0013. We are grateful to Z. Ammari, M. Lemm, J. Zhang and especially I.M. Sigal for fruitful collaborations. 

\section{Notations and Definitions}\label{sec:Notations-Definitions}

We recall that $d$ is the dimension, we will thus always assume that $d$ is in $\mathbb{N}$. The distance between two subsets $X$ and $Y$ of $\mathbb{R}^d$ is
\begin{equation*}
    \dXY:=\inf\{|x-y|\mid x\in X, y\in Y\}.
\end{equation*}
For $\xi$ in $\mathbb{R}^d$, $\langle\xi\rangle:=\sqrt{1+|\xi|^2}$ and likewise $\jnabla := \sqrt{1-\Delta}$.

The functional spaces below are spaces of functions from $\mathbb{R}^d$ to $\mathbb{C}$. The Banach space of (equivalent classes of) Lebesgue square integrable functions is denoted by $L^2$. The set of compactly supported smooth functions is denoted by $\mathcal{C}^\infty_0$. The Schwartz class is denoted by~$\mathcal{S}$. The Fourier transform of a tempered distribution $\psi$ in the dual of the Schwartz class~$\mathcal{S}'$ is denoted by $\mathcal{F}(\psi)$ (recall that we omit the choice of the normalization of the Fourier transform as it is irrelevant for our purpose). For $s$ in $\mathbb{R}$, $H^{s}$ is the usual Sobolev space, 
\begin{equation*}
    H^s:=\{\psi\in \mathcal{S}'\mid \mathcal{F}(\psi)\in L^1_{\mathrm{loc}} \text{ and } \xi \mapsto \langle\xi\rangle^s\mathcal{F}(\psi)(\xi)\in L^2\}
\end{equation*}
and $\dot{H}^{s}$ is the corresponding  homogeneous Sobolev space, 
\begin{equation*}
    \dot{H}^s:=\{ \psi\in \mathcal{S}'\mid \mathcal{F}(\psi)\in L^1_{\mathrm{loc}} \text{ and } \xi \mapsto |\xi|^s\mathcal{F}(\psi)(\xi)\in L^2\}.
\end{equation*}
The set of norm continuous operators from a Banach space $\mathcal{V}_1$ to a Banach space $\mathcal{V}_2$ is denoted by $\mathcal{B}(\mathcal{V}_1,\mathcal{V}_2)$. If $\mathcal{V}_1=\mathcal{V}_2$ we set $\mathcal{B}(\mathcal{V}_1):=\mathcal{B}(\mathcal{V}_1,\mathcal{V}_1)$. The domain and quadratic form domain of an operator $A$ in a Hilbert space are denoted by $\mathcal{D}(A)$ and $\mathcal{Q}(A)$, respectively.

We now specify the notion of unitary propagator that we use here.
 We consider a compact interval  $I$ of $\mathbb{R}$ and a family $(A_t)_{t\in I}$ of self-adjoint operators on $L^2$ such that $\mathcal{D}(A_t)\cap H^{1/2}$ is dense in $H^{1/2}$ and the $A_t$ are continuously extendable to $\mathcal{B}(H^{1/2},H^{-1/2})$.

\begin{definition}\label{def_propagator}
The map $I\times I \ni (t,s)\mapsto U(t,s)$ is a \emph{unitary propagator} associated to
\begin{equation}
                i\partial_t \psi_t = A_t \psi_t\,,  t\in I
 \label{Cauchy-pb-propagateur}
\end{equation}
if and only if
\begin{enumerate}
    \item $U(t,s)$ is unitary on $L^2$ for all $t,s$ in $I$,
    \item $U(t,t)=\mathbf{1}_{L^2}$ for all $t$ in $I$ and $U(t,s)U(s,r)=U(t,r)$ for all $t,s,r$ in $I$,
    \item For all $s$ in $I$, the map $t\ni I \mapsto U(t,s)$ belongs to 
    \begin{equation*}
        \mathcal{C}^0(I,\mathcal{B}(H^{\frac12})_{\mathrm{str}}) \cap \mathcal{C}^1(I,\mathcal{B}(H^{\frac12},H^{-\frac12})_{\mathrm{str}})
    \end{equation*}
    and satisfies
    \[\forall t,s\in I\,,\,\forall \psi \in H^{\frac12}\,, \quad i\partial_t U(t,s)\psi = A_t U(t,s)\psi\,,\]
    as an equality in $H^{-1/2}$.
\end{enumerate}
\end{definition}
In the previous definition, the index ``$\mathrm{str}$'' indicates that the considered topology is the strong operator topology.
In the sequel we will apply Definition \ref{def_propagator} with $I=[0,T]$. As mentioned in the introduction, we will use the notation $U_t=U(t,0)$.

\begin{rem}
    It is not difficult to verify that if $U(t,s)$ is a unitary propagator in the sense of Definition \ref{def_propagator}, then we also have
    \begin{equation*}
        \forall t,s\in I\,,\,\forall \psi \in H^{\frac12}\,, \quad i\partial_s U(t,s)\psi = U(t,s)A_s\psi\,.
    \end{equation*}
\end{rem}

\section{Maximal Velocity Estimates}\label{sec:max-vel-estimates}
In this section we prove Theorem \ref{th:max-vel} and Corollary \ref{cor:max-vel-bounds-general-subsets}. Note that if $X$ and $Y$ are such that $\mathrm{dist}(X,Y)=0$, then the statements of Theorem \ref{th:max-vel} and Corollary \ref{cor:max-vel-bounds-general-subsets} are obvious. In the remainder of this section we will therefore assume that $\mathrm{dist}(X,Y)>0$.

Following the strategy explained in the introduction we aim at proving \eqref{eq:idee-borne-de-vitesse-maximale} for a suitable function $\ell$.
To reach those estimates, we will need several lemmata. The first one is a quantitative separation lemma which allows us to introduce the function $\ell$ satisfying Eq.~\eqref{eq:idee-borne-de-vitesse-maximale}.
	\begin{lemma}\label{lm:convex}
		Let $X,Y$ be two convex subsets of~$\mathbb{R}^d$ such that~$\dXY>0$. 
		There exist $x_0$ in $\mathbb{R}^d$ and a unit vector $n$  in $\mathbb{R}^d$ such that the affine functional~$\ell(x):=n\cdot (x-x_0)$ satisfies
		\begin{equation*}
		    \forall x\in X,\quad \ell(x)\ge\frac12 \dXY \qquad \text{and} \qquad \forall x\in Y,\quad \ell(x)\le-\frac12 \dXY\,.
		\end{equation*}
\end{lemma}

	\begin{proof}
		We use the notation $\mathring{B}(a,r)$ for the open ball centered at $a$ and of radius $r>0$.  The sets
		\[A_X:=\bigcup_{x\in {X}} \mathring{B}\Big(x,\frac{\dXY}{2}\Big)\quad \text{and}\quad A_Y:=\bigcup_{x\in {Y}} \mathring{B}\Big(x,\frac{\dXY}{2}\Big)
		\]
		are convex and disjoint, and one can use the separation of disjoint convex sets to get $n,x_0$ in~$\mathbb{R}^d$, $|n|=1$ such that $\ell(x)=n\cdot (x-x_0)$ is nonnegative on $A_X$ and nonpositive on $A_Y$. By continuity, $\ell$ is nonnegative on $\overline{A_X}$ and for all $x$ in $X$, $x-\frac{\dXY}{2}n$ is in~$\overline{A_X}$ and thus\[ \ell(x)-\frac{\dXY}{2}=n\cdot\Big(x-\frac{\dXY}{2}n-x_0\Big)\geq 0\,.\]
		The inequality for $x$ in $Y$ is obtained using that $x+\frac{\dXY}{2}n$ lies in~$\overline{A_Y}$.
	\end{proof}
	
	From now on, we consider $\ell$ as in Lemma~\ref{lm:convex} and for all $\varepsilon>0$, we introduce a bounded regularization of $\ell$ by setting
	    \begin{equation*}
      \ell_\varepsilon(x):=f_\varepsilon(\ell(x))=f_\varepsilon(n\cdot(x-x_0)),
	    \end{equation*}
	    where $f_\varepsilon(r)=f(\varepsilon r)$, $f$ belongs to $C^\infty(\mathbb{R})$, $f(r)=r$ on $[-1,1]$, $0\le f'\le1$ and $f'$ is compactly supported. 
	    We also introduce the notation
	    \begin{equation*}
	        \nabla_{\pm n,\varepsilon}:=\nabla\pm n f'_\varepsilon(\ell(x)), \qquad \Delta_{\pm n,\varepsilon}:=\big(\nabla\pm n f'_\varepsilon(\ell(x))\big)^2.
	    \end{equation*}
We recall (see \cite{ReedSimon1}) that a quadratic form $Q$ on $L^2$ with form domain $\mathcal{Q}$ is called strictly $m$-accretive if it closed on $\mathcal{Q}$ and there exists $0<\theta<\frac\pi2$ such that $|\mathrm{Arg}\, Q(\varphi,\varphi)|\le\theta$ for all $\varphi\in\mathcal{Q}$. By \cite[Theorem VIII.16]{ReedSimon1}, if $Q$ is strictly $m$-accretive, there is a unique closed operator $A$ associated to $Q$, and for all $\lambda>0$, we have
\begin{equation*}
    \|(A+\lambda)^{-1}\|_{\mathcal{B}(L^2)}\le\lambda^{-1}.
\end{equation*}

	\begin{lemma}
	    For all $\varepsilon\ge0$ and $\zeta>0$, $-\Delta_{\pm n,\varepsilon}$ is strictly $m$-accretive on $H^1$ and we have
	\begin{align}
    & \big\|\big[-\Delta_{\pm n,\varepsilon}+1+\zeta \big]^{-1}\big\|_{\mathcal{B}(L^2)}\le \zeta^{-1},\label{eq:bound-res1}\\
    & \big\|\nabla\big[-\Delta_{\pm n,\varepsilon}+1+\zeta \big]^{-1}\big\|_{\mathcal{B}(L^2)}\lesssim \max(\zeta^{-\frac12},\zeta^{-1}),\label{eq:bound-res2}\\
    & \big\|\nabla\big[-\Delta_{\pm n,\varepsilon}+1+\zeta\big]^{-1}\nabla\big\|_{\mathcal{B}(L^2)}\lesssim \max(1,\zeta^{-1}).\label{eq:bound-res3}
\end{align}
	\end{lemma}

	\begin{proof}
Let $\varepsilon\ge0$, $\zeta>0$. We compute, for all $\varphi$ and $\psi$ in $H^1$, 
\begin{multline*}
    \big\langle\varphi,\big[-\Delta_{\pm n,\varepsilon}+1+\zeta\big]\psi\big\rangle\\
    =\big\langle\varphi,\big[-\Delta+1+\zeta-f'_\varepsilon(\ell(x))^2\big]\psi\big\rangle\mp\big\langle\varphi,\big[\nabla\cdot n f'_\varepsilon(\ell(x))+f'_\varepsilon(\ell(x))n\cdot\nabla\big]\psi\big\rangle.
\end{multline*}
Since $0\le f'_\varepsilon\le1$, this implies that
\begin{align*}
    \big|\big\langle\varphi,\big[-\Delta_{\pm n,\varepsilon}+1+\zeta\big]\psi\big\rangle\big|\lesssim\|\varphi\|_{H^1}\|\psi\|_{H^1},
\end{align*}
and hence $-\Delta_{\pm n,\varepsilon}+1+\zeta$ is a well-defined quadratic form on $H^1$. Now,
\begin{align*}
    \mathrm{Re} \, \big\langle\varphi,\big[-\Delta_{\pm n,\varepsilon}+1+\zeta\big]\varphi\big\rangle&=\big\langle\varphi,\big[-\Delta+1+\zeta-f'_\varepsilon(\ell(x))^2\big]\varphi\big\rangle\ge\|\varphi\|_{\dot{H}^1}^2+\zeta\|\varphi\|_{L^2}^2,
\end{align*}
and
\begin{align*}
    \big|\mathrm{Im} \, \big\langle\varphi,\big[-\Delta_{\pm n,\varepsilon}+1+\zeta\big]\varphi\big\rangle\big|&=\big|\big\langle\varphi,\big[\nabla\cdot n f'_\varepsilon(\ell(x))+f'_\varepsilon(\ell(x))n\cdot\nabla\big]\varphi\big\rangle\big|\le2\|\varphi\|_{\dot{H}^1}\|\varphi\|_{L^2}.
\end{align*}
The previous two equations imply that the quadratic form $-\Delta_{\pm n,\varepsilon}+1+\zeta$ is closed on $H^1$. Moreover, for all $\varphi$ in $H^1$,
\begin{align*}
    \big\langle\varphi,\big[-\Delta_{\pm n,\varepsilon}+1+\zeta\big]\psi\big\rangle\in\big\{z=\lambda+i\mu\in\mathbb{C}\,|\,\lambda\ge0,\, |\mu|\le \zeta^{-\frac12}\lambda\big\},
\end{align*}
and hence $-\Delta_{\pm n,\varepsilon}+1+\zeta$ is strictly $m$-accretive. By \cite[Theorem VIII.17]{ReedSimon1}, we then deduce that 
for all $\zeta'>0$,
\begin{align*}
    \big\|\big[-\Delta_{\pm n,\varepsilon}+1+\zeta+\zeta'\big]^{-1}\big\|_{\mathcal{B}(L^2)}\le (\zeta')^{-1}.
\end{align*}
Since this holds for any $\zeta>0$, \eqref{eq:bound-res1} follows.

To prove \eqref{eq:bound-res2}, it suffices to write, for any $\varphi$ in $L^2$,
\begin{align*}
    \big\|\nabla&\big[-\Delta_{\pm n,\varepsilon}+1+\zeta\big]^{-1}\varphi\big\|^2_{L^2}\\
    &=-\big\langle\big[-\Delta_{\pm n,\varepsilon}+1+\zeta\big]^{-1}\varphi,\Delta\big[-\Delta_{\pm n,\varepsilon}+1+\zeta\big]^{-1}\varphi\big\rangle\\
    &=-\big\langle\big[-\Delta_{\pm n,\varepsilon}+1+\zeta\big]^{-1}\varphi,\Delta_{\pm n,\varepsilon}\big[-\Delta_{\pm n,\varepsilon}+1+\zeta\big]^{-1}\varphi\big\rangle+\mathrm{Rem}.
\end{align*}
By \eqref{eq:bound-res1}, the first term is bounded by $\mathcal{O}(\zeta^{-1})\|\varphi\|^2_{L^2}$  while the ``remainder'' term satisfies \begin{equation*}
    \|\mathrm{Rem}\|\lesssim \zeta^{-1}\big\|\nabla\big[-\Delta_{\pm n,\varepsilon}+1+\zeta\big]^{-1}\varphi\big\|_{L^2}\|\varphi\|_{L^2} + \|\big[-\Delta_{\pm n,\varepsilon}+1+\zeta\big]^{-1}\varphi\big\|^2_{L^2}.
\end{equation*}
This yields \eqref{eq:bound-res2}. To prove the last equation \eqref{eq:bound-res3}, we proceed similarly, first writing the equation $\nabla=\nabla\pm n f'_\varepsilon(\ell(x))\mp n f'_\varepsilon(\ell(x))$ and then using that $\nabla\pm n f'_\varepsilon(\ell(x))$ commutes with the resolvent.
	\end{proof}
	
	For all $\varepsilon>0$, we define the operator $\Teps$ on $H^1$ by
	    \begin{equation}\label{eq:defTeps}
	        \Teps:=\mathrm{Im}\big(e^{\ell_\varepsilon(x)} \langle\nabla\rangle e^{-\ell_\varepsilon(x)}\big)=\frac{1}{2i}\big(e^{\ell_\varepsilon(x)} \langle\nabla\rangle e^{-\ell_\varepsilon(x)}-e^{-\ell_\varepsilon(x)} \langle\nabla\rangle e^{\ell_\varepsilon(x)}\big).
	    \end{equation}

	\begin{lemma}\label{lm:unif-bound}
	    For all $\varepsilon>0$, $\Teps$ extends to a bounded operator on $L^2$, with
	    \begin{equation*}
	        \sup_{\varepsilon>0}\|\Teps\|_{\mathcal{B}(L^2)}<\infty.
	    \end{equation*}
	\end{lemma}
	
	\begin{proof}
	   Using 
	   the relation
\begin{equation*}
			\langle\nabla\rangle = \frac{1}{\pi} \int_0^\infty\left( \zeta^{-1/2} - \zeta^{1/2}(-\Delta+1+\zeta)^{-1} \right)\, \mathrm{d}\zeta,
		\end{equation*}
on $H^1$ (which directly follows from functional calculus), we have, as a quadratic form on $H^1$,
\begin{multline}
   e^{\ell_\varepsilon(x)} \langle\nabla\rangle e^{-\ell_\varepsilon(x)}-e^{-\ell_\varepsilon(x)} \langle\nabla\rangle e^{\ell_\varepsilon(x)} \\
   =\frac{1}{\pi} \int_0^\infty \zeta^{1/2}\Big[\big(-\Delta_{n,\varepsilon}+1+\zeta\big)^{-1}-\big(-\Delta_{- n,\varepsilon}+1+\zeta\big)^{-1}\Big]\, \mathrm{d}\zeta. \label{eq:int-imaginary-part}
\end{multline}
Here we used the explicit computation $e^{\pm \ell_\varepsilon(x)} \nabla e^{\mp \ell_\varepsilon(x)} = \nabla \mp n f'_\varepsilon(\ell(x)) = \nabla_{\mp\, n,\varepsilon}$.

Now we split the integral in the right-hand side of \eqref{eq:int-imaginary-part} into two parts. For the integral from $0$ to $1$, using \eqref{eq:bound-res1} we directly obtain that
\begin{align}
    \int_0^1 \zeta^{1/2}\Big\|\big[-\Delta_{n,\varepsilon}+1+\zeta\big]^{-1}-\big[-\Delta_{- n,\varepsilon}+1+\zeta\big]^{-1}\Big\|_{\mathcal{B}(L^2)}\, \mathrm{d}\zeta \lesssim 1,
\end{align}
uniformly in $\varepsilon>0$. For the integral from $1$ to $\infty$, we use first the resolvent equation, writing
\begin{align}
    &\big[-\Delta_{ n,\varepsilon}+1+\zeta\big]^{-1}-\big[-\Delta_{- n,\varepsilon}+1+\zeta\big]^{-1}\nonumber\\
    &=2\big[-\Delta_{- n,\varepsilon}+1+\zeta\big]^{-1}\big[\nabla\cdot n f'_\varepsilon(\ell(x))+f'_\varepsilon(\ell(x))n\cdot\nabla\big]\big[-\Delta_{ n,\varepsilon}+1+\zeta\big]^{-1}\nonumber\\
    &=2\nabla_{-n,\varepsilon} \cdot n \big[-\Delta_{- n,\varepsilon}+1+\zeta\big]^{-1}f'_\varepsilon(\ell(x))\big[-\Delta_{n,\varepsilon}+1+\zeta\big]^{-1}\nonumber\\
    &\quad +2 \big[-\Delta_{- n,\varepsilon}+1+\zeta\big]^{-1}f'_\varepsilon(\ell(x))\big[-\Delta_{ n,\varepsilon}+1+\zeta\big]^{-1}n\cdot\nabla_{ n,\varepsilon}. \label{eq:once-resolv}
\end{align}
The two terms on the right-hand side are estimated in the same way. Consider for instance the first one. Using again the resolvent equation, we obtain
\begin{align}
    &\nabla_{- n,\varepsilon}\cdot n \big[-\Delta_{- n,\varepsilon}+1+\zeta\big]^{-1}f'_\varepsilon(\ell(x))\big[-\Delta_{ n,\varepsilon}+1+\zeta\big]^{-1}\notag\\
    &=\nabla_{- n,\varepsilon}\cdot n \big[-\Delta+1+\zeta\big]^{-1}f'_\varepsilon(\ell(x))\big[-\Delta+1+\zeta\big]^{-1}\notag\\
    &\quad+\nabla_{- n,\varepsilon}\cdot n\big[-\Delta+1+\zeta\big]^{-1}\big(f'_\varepsilon(\ell(x))^2+f'_\varepsilon(\ell(x))n\cdot\nabla+\nabla\cdot n f'_\varepsilon(\ell(x))\big)\notag\\
    &\qquad\qquad\big[-\Delta_{- n,\varepsilon}+1+\zeta\big]^{-1}f'_\varepsilon(\ell(x))\big[-\Delta_{ n,\varepsilon}+1+\zeta\big]^{-1}\notag\\
        &\quad+\nabla_{- n,\varepsilon}\cdot n\big[-\Delta+1+\zeta\big]^{-1}f'_\varepsilon(\ell(x))\big[-\Delta+1+\zeta\big]^{-1}\notag\\
    &\qquad\qquad\big(f'_\varepsilon(\ell(x))^2+f'_\varepsilon(\ell(x))n\cdot\nabla+\nabla\cdot n f'_\varepsilon(\ell(x))\big)\big[-\Delta_{ n,\varepsilon}+1+\zeta\big]^{-1}.\label{eq:twice-resolv}
\end{align}
Using in particular \eqref{eq:bound-res1}--\eqref{eq:bound-res3}, it is not difficult to see that the last two terms in the right-hand side of~\eqref{eq:twice-resolv} are $\mathcal{O}(\zeta^{-2})$, uniformly in $\varepsilon$. For the first term in the right-hand side of~\eqref{eq:twice-resolv}, we commute $f'_\varepsilon(\ell(x))$ through $\big[-\Delta+1+\zeta\big]^{-1}$, obtaining
\begin{align*}
    &\nabla_{- n,\varepsilon}\cdot n \big[-\Delta+1+\zeta\big]^{-1}f'_\varepsilon(\ell(x))\big[-\Delta+1+\zeta\big]^{-1}\\
    &=\nabla_{- n,\varepsilon}\cdot n \big[-\Delta+1+\zeta\big]^{-2}f'_\varepsilon(\ell(x))\\
    &\quad+\nabla_{- n,\varepsilon}\cdot n \big[-\Delta+1+\zeta\big]^{-2}\big(\nabla\cdot(\nabla f'_\varepsilon)(\ell(x))+(\nabla f'_\varepsilon)(\ell(x))\cdot\nabla\big)\big[-\Delta+1+\zeta\big]^{-1}.
\end{align*}
The second term is $\mathcal{O}(\zeta^{-2})$ by the same arguments as before. Therefore, combining the previous expression with \eqref{eq:twice-resolv}, we have  established that
\begin{align*}
    &\int_1^\infty \zeta^{1/2}\Big(\big[-\Delta_{- n,\varepsilon}+1+\zeta\big]^{-1}-\big[-\Delta_{ n,\varepsilon}+1+\zeta\big]^{-1}\Big)\, \mathrm{d}\zeta\\
    &=\int_1^\infty \zeta^{1/2}\Big( \nabla_{- n,\varepsilon}\cdot n \big[-\Delta+1+\zeta\big]^{-2}f'_\varepsilon(\ell(x))-  f'_\varepsilon(\ell(x))\big[-\Delta+1+\zeta\big]^{-2}\nabla_{ n,\varepsilon}\cdot n\Big)\, \mathrm{d}\zeta +\mathrm{R},
\end{align*}
with $\|\mathrm{R}\|_{\mathcal{B}(L^2)}\lesssim1$ uniformly in $\varepsilon$. Here the second term in the integral is the contribution from the term in \eqref{eq:once-resolv}. Now we can replace the integral from $1$ to $\infty$ by the integral from~$0$ to $\infty$ up to a uniformly bounded contribution, and then integrate in $\zeta$, using the explicit expression
\begin{equation*}
    \langle \nabla \rangle ^{-1} = \frac{2}{\pi} \int_0^\infty \zeta^{1/2} (-\D + 1+\zeta)^{-2}\, \mathrm{d}\zeta\,.
\end{equation*}
Since
 \begin{align*}
     \big\|\nabla_{- n,\varepsilon}\cdot n \langle\nabla\rangle^{-1} f'_\varepsilon(\ell(x))\big\|_{\mathcal{B}(L^2)}\lesssim1, \qquad \big\| f'_\varepsilon(\ell(x)) \langle\nabla\rangle^{-1}\nabla_{ n,\varepsilon}\cdot n\big\|_{\mathcal{B}(L^2)}\lesssim1,
 \end{align*}
 uniformly in $\varepsilon$, this concludes the proof of the lemma.
	\end{proof}

For all $z$ in $\mathbb{C}\setminus(-\infty,0)$, we write $\sqrt{z}=\sqrt{|z|}e^{\frac{i}{2}\mathrm{Arg}(z)}$ with $-\pi<\mathrm{Arg}(z)<\pi$ and for all~$\xi$ in $\mathbb{R}^d$, we set
	\begin{align}\label{eq:def_fpm}
    f_\pm(\xi):=\sqrt{|\xi\pm in|^2+1}=\sqrt{|\xi|^2\pm 2in\cdot\xi}.
\end{align}

\begin{lemma}\label{lm:bound_fpm}
    For all $\xi$ in $\mathbb{R}^d$, we have
    \begin{equation*}
        \big|\mathrm{Im}f_\pm(\xi)\big|\le1.
    \end{equation*}
\end{lemma}

\begin{proof}
    A direct computation shows that, for all $z=\lambda+i\mu$ with $\lambda,\mu$ in $\mathbb{R}$,
    \begin{align*}
        \mathrm{Im}\sqrt{z}=\frac{\mathrm{sign}(\mu)}{\sqrt{2}}\big(\sqrt{\lambda^2+\mu^2}-\lambda\big)^\frac12.
    \end{align*}

Applying this with $\lambda=|\xi|^2$, $\mu=\pm2 n\cdot\xi$, we obtain
\begin{align*}
    |\mathrm{Im}f_\pm(\xi)|=\frac1{\sqrt{2}}\big(\sqrt{|\xi|^4+4(n\cdot\xi)^2}-|\xi|^2\big)^\frac12=\sqrt{2}\frac{|n\cdot\xi|}{\big(\sqrt{|\xi|^4+4(n\cdot\xi)^2}+|\xi|^2\big)^\frac12}.
\end{align*}
Since $|n\cdot\xi|^2\le|\xi|^2\le\frac12(\sqrt{|\xi|^4+4(n\cdot\xi)^2}+|\xi|^2)$, the result follows.
\end{proof}

We define the operator $\Tzero$ on $L^2$ by
\begin{equation}\label{eq:defT0}
    \Tzero := \mathrm{Im}(f_+(-i\nabla)) = \mathcal{F} \, \mathrm{Im}(f_+(\xi)) \,  \mathcal{F}^{-1}.
\end{equation}
It then follows from Lemma \ref{lm:bound_fpm} that
\begin{equation}
    \|\Tzero\|_{\mathcal{B}(L^2)}\le1. \label{eq:borne-T0}
\end{equation}
The next lemma shows that $\Tzero$ is the weak limit of $\Teps$ (defined in \eqref{eq:defTeps}) as $\varepsilon\to0$.

\begin{lemma}\label{lm:strongconv}
We have
\begin{equation*}
    \Teps \to \Tzero, \quad \varepsilon\to0,
\end{equation*}
weakly in $\mathcal{B}(L^2)$.
\end{lemma}

\begin{proof}
We first show that for all $\varphi$ and $\psi$ in $\mathcal{C}_0^\infty$,
\begin{equation}\label{eq:identity_T0}
    \langle\varphi,\Tzero\psi\rangle=\mathrm{Im}\big\langle e^{\ell(x)}\varphi,\langle\nabla\rangle e^{-\ell(x)}\psi\big\rangle=\mathrm{Im}\big\langle e^{n\cdot x}\varphi,\langle\nabla\rangle e^{-n\cdot x}\psi\big\rangle.
\end{equation}
The second equality is obvious. To prove the first one, let us set $\tilde\varphi(x)=e^{n\cdot x}\varphi(x)$ and, similarly, $\tilde \psi(x)=e^{-n\cdot x}\psi(x)$. We have $\varphi,\tilde\varphi,\psi,\tilde\psi$ in $\mathcal{C}_0^\infty$ and hence the Paley-Wiener Theorem (see e.g. \cite[Theorem IX.11]{reed_sim_vol2}) implies that the Fourier transforms of these functions are entire analytic on $\mathbb{C}^d$, satisfying, for all integer $j$ in $\mathbb{N}$,
\begin{equation}\label{eq:PaleyWiener}
    \big |(\mathcal{F}\varphi)(z)\big|\le C_n(1+|z|)^{-j}e^{R|\mathrm{Im}(z)|}, \quad z\in\mathbb{C}^d,
\end{equation}
for some $R>0$, and likewise for $\tilde\varphi$, $\psi$, $\tilde\psi$. Since in addition $\mathcal{F}(\tilde\varphi)(z)=\mathcal{F}(\varphi)(z+in)$ and, similarly, $\mathcal{F}(\tilde\psi)(z)=\mathcal{F}(\psi)(z-in)$, we can compute
\begin{align*}
    \big\langle e^{n\cdot x}\varphi,\langle\nabla\rangle e^{-n\cdot x}\psi\big\rangle&=\int_{\mathbb{R}^d}\overline{\mathcal{F}(\varphi)(\xi+in)}\,\langle\xi\rangle\,\mathcal{F}(\psi)(\xi-in)\diff\xi\\
    &=\int_{\mathbb{R}^d}\mathcal{F}^{-1}(\bar\varphi)(\xi-in)\,\langle\xi\rangle\,\mathcal{F}(\psi)(\xi-in)\diff\xi.
\end{align*}
Using analyticity and the decay properties \eqref{eq:PaleyWiener}, we can shift the contour of integration in the previous integral, obtaining
\begin{align*}
    \big\langle e^{n\cdot x}\varphi,\langle\nabla\rangle e^{-n\cdot x}\psi\big\rangle
    &=\int_{\mathbb{R}^d}\mathcal{F}^{-1}(\bar\varphi)(\xi)\,\langle\xi+in\rangle\,\mathcal{F}(\psi)(\xi)\diff\xi\\
    &=\int_{\mathbb{R}^d}\overline{\mathcal{F}(\varphi)(\xi)}\,\langle\xi+in\rangle\,\mathcal{F}(\psi)(\xi)\diff\xi=\big\langle\varphi,f_+(-i\nabla)\psi\big\rangle.
\end{align*}
Taking the imaginary part gives \eqref{eq:identity_T0}.

Now we prove the weak convergence in the statement of the lemma. Let $\varphi,\psi$ in $ L^2$. Let~$0<\delta<1$ and $\varphi_\delta,\psi_\delta$ in $\mathcal{C}_0^\infty$ be such that $\|\varphi-\varphi_\delta\|_{L^2}\le\delta$, $\|\psi-\psi_\delta\|_{L^2}\le\delta$. Using Lemma~\ref{lm:unif-bound} and $\|\Tzero\|_{\mathcal{B}(L^2)}\le1$, we have
\begin{align}
   &\big|\langle\varphi,\Teps\psi\rangle-\langle\varphi,\Tzero\psi\rangle\big|\le\big|\langle\varphi_\delta,\Teps\psi_\delta\rangle-\langle\varphi_\delta,\Tzero\psi_\delta\rangle\big|+C\delta,\label{eq:weakconv0}
\end{align}
uniformly in $\varepsilon>0$. Moreover, we can also write
\begin{align*}
    &\langle e^{\pm\ell_\varepsilon(x)}\varphi_\delta, \langle\nabla\rangle e^{\mp\ell_\varepsilon(x)}\psi_\delta\rangle-\langle e^{\pm\ell(x)}\varphi_\delta, \langle\nabla\rangle e^{\mp\ell(x)}\psi_\delta\rangle\\
    &=\big\langle\big(e^{\pm\ell_\varepsilon(x)}-e^{\pm\ell(x)}\big)\varphi_\delta, \langle\nabla\rangle e^{\mp\ell_\varepsilon(x)}\psi_\delta\big\rangle+\big\langle \langle\nabla\rangle e^{\pm\ell(x)}\varphi_\delta,  \big( e^{\mp\ell_\varepsilon(x)}-e^{\mp\ell(x)}\big)\psi_\delta\big\rangle.
\end{align*}
Since $\varphi_\delta,\psi_\delta\in\mathcal{C}_0^\infty$, we have $\|\langle\nabla\rangle e^{\pm\ell(x)}\varphi_\delta\|_{L^2}\le C_\delta$, $\|\langle\nabla\rangle e^{\mp\ell_\varepsilon(x)}\psi_\delta\|_{L^2}\le C_\delta$ (uniformly in $\varepsilon$), and
\begin{equation*}
    \big\|\big(e^{\pm\ell_\varepsilon(x)}-e^{\pm\ell(x)}\big)\varphi_\delta\big\|_{L^2}\to0, \quad \big\|\big(e^{\mp\ell_\varepsilon(x)}-e^{\mp\ell(x)}\big)\psi_\delta\big\|_{L^2}\to0, \quad\varepsilon\to0.
\end{equation*}
Now, using \eqref{eq:identity_T0} we can insert this into \eqref{eq:weakconv0}, which yields
\begin{equation*}
    \limsup_{\varepsilon\to0}\big|\langle\varphi,\Teps\psi\rangle-\langle\varphi,\Tzero\psi\rangle\big|\le C\delta.
\end{equation*}
Since $\delta>0$ is arbitrary, this concludes the proof.
\end{proof}

	We now consider the unitary propagator $(U_t)_{t\in[0,T]}$ generated by $(\jnabla+V_t)_{t\in[0,T]}$ as in the statement of Theorem \ref{th:max-vel}. We first use Lemma \ref{lm:unif-bound} to show that for any $t$ in $[0,T]$, the operator $e^{-\ell(x)}U_te^{\ell(x)}$ is well-defined and bounded on $L^2$.
	
	\begin{lemma}\label{lm:bound_domain}Under the assumptions of Theorem~\ref{th:max-vel}, for all $t$ in $[0,T]$, we have
    \begin{equation*}
        \mathrm{Ran}(U_te^{\ell(x)})\subset\mathcal{D}(e^{-\ell(x)}),
    \end{equation*}
    and $e^{-\ell(x)}U_te^{\ell(x)}$ extends to a bounded operator on $L^2$. Moreover,
    \begin{equation*}
        e^{-\ell_\varepsilon(x)}U_te^{\ell_\varepsilon(x)} \to e^{-\ell(x)}U_te^{\ell(x)} \quad \text{as}\quad \varepsilon\to0,
    \end{equation*}
    strongly in $\mathcal{B}(L^2)$. 
\end{lemma}
	
	\begin{proof}
	    Let $\varphi$ in $\mathcal{C}^\infty_0$ and  $\varepsilon>0$.  We  compute
	    \begin{align*}
    &\big\|e^{-\ell_\varepsilon(x)}U_te^{\ell_\varepsilon(x)}\varphi\big\|_{L^2}^2\\
    &=\|\varphi\|_{L^2}^2+2\mathrm{Re}\int_0^t\big\langle e^{-\ell_\varepsilon(x)}U_\tau e^{\ell_\varepsilon(x)}\varphi \, , \, e^{-\ell_\varepsilon(x)} (-iH_\tau)U_\tau e^{\ell_\varepsilon(x)}\varphi\big\rangle\mathrm{d}\tau\\
    &=\|\varphi\|_{L^2}^2+2\mathrm{Re}\int_0^t\big\langle e^{-\ell_\varepsilon(x)}U_\tau e^{\ell_\varepsilon(x)}\varphi \, , \, e^{-\ell_\varepsilon(x)} (-i\langle\nabla\rangle)e^{\ell_\varepsilon(x)}e^{-\ell_\varepsilon(x)}U_\tau e^{\ell_\varepsilon(x)}\varphi\big\rangle\mathrm{d}\tau,
\end{align*}
where in the second equality we used that
\begin{equation*}
    \mathrm{Re}\big\langle e^{-\ell_\varepsilon(x)}U_\tau e^{\ell_\varepsilon(x)}\varphi \, , \, e^{-\ell_\varepsilon(x)} (-iV_t)e^{\ell_\varepsilon(x)}e^{-\ell_\varepsilon(x)}U_\tau e^{\ell_\varepsilon(x)}\varphi\big\rangle=0.
\end{equation*}
Since
\begin{align*}
    2\mathrm{Re}\, e^{-\ell_\varepsilon(x)} (-i\langle\nabla\rangle)e^{\ell_\varepsilon(x)}&=2\mathrm{Im}\,\big( e^{-\ell_\varepsilon(x)}\langle\nabla\rangle e^{\ell_\varepsilon(x)}\big)=2\Teps, 
\end{align*}
we can rewrite the equality above as
	\begin{align}\label{eq:Gronwall}
    &\big\|e^{-\ell_\varepsilon(x)}U_te^{\ell_\varepsilon(x)}\varphi\big\|_{L^2}^2=\|\varphi\|_{L^2}^2+2\int_0^t\big\langle e^{-\ell_\varepsilon(x)}U_\tau e^{\ell_\varepsilon(x)}\varphi \, , \, \Teps e^{-\ell_\varepsilon(x)}U_\tau e^{\ell_\varepsilon(x)}\varphi\big\rangle\mathrm{d}\tau.
\end{align}
Now we can use Lemma \ref{lm:unif-bound} to deduce that
\begin{align*}
    &\big\|e^{-\ell_\varepsilon(x)}U_te^{\ell_\varepsilon(x)}\varphi\big\|_{L^2}^2
    \le\|\varphi\|_{L^2}^2+C\int_0^t\big\| e^{-\ell_\varepsilon(x)}U_\tau e^{\ell_\varepsilon(x)}\varphi\big\|_{L^2}^2\mathrm{d}\tau,
\end{align*}
for some positive constant $C$ independent of $\varepsilon$. Hence, by Gronwall's inequality,
\begin{align}\label{eq:Gronwall-eps}
    &\big\|e^{-\ell_\varepsilon(x)}U_te^{\ell_\varepsilon(x)}\varphi\big\|_{L^2}^2
    \le e^{Ct}\|\varphi\|_{L^2}^2,
\end{align}
uniformly in $\varepsilon$. By density, this inequality can be extended to any $\varphi$ in $L^2$.

For all $\varphi$ in $\mathcal{D}(e^{\ell(x)})$ and $\psi$ in $\mathcal{D}(e^{-\ell(x)})$, using that $\|U_t\|_{\mathcal{B}(L^2)}=1$, we can then write
	    \begin{align*}
	        \big|\langle e^{-\ell(x)}\psi , U_te^{\ell(x)}\varphi\rangle\big|
	        &=\lim_{\varepsilon\to0}\big|\langle \psi , e^{-\ell_\varepsilon(x)}U_te^{\ell_\varepsilon(x)}\varphi\rangle\big|
	        \le e^{\frac12Ct}\|\psi\|_{L^2}\|\varphi\|_{L^2}.
	   \end{align*}
This proves that $\mathrm{Ran}(U_te^{\ell(x)})\subset\mathcal{D}(e^{-\ell(x)})$ and that $e^{-\ell(x)}U_te^{\ell(x)}$ extends to a bounded operator on $L^2$.

To prove the strong convergence, consider now $\varphi$ in $L^2$. Let $\delta>0$ and let $\varphi_\delta$ in $\mathcal{C}_0^\infty$ be such that $\|\varphi-\varphi_\delta\|_{L^2}\le\delta$. We write
\begin{align}
   & e^{-\ell(x)}U_te^{\ell(x)}\varphi-e^{-\ell_\varepsilon(x)}U_te^{\ell_\varepsilon(x)}\varphi\notag \\
    & = (e^{-\ell(x)}-e^{-\ell_\varepsilon(x)})U_te^{\ell(x)}\varphi_\delta-e^{-\ell_\varepsilon(x)}U_te^{\ell(x)}(e^{\ell_\varepsilon(x)}e^{-\ell(x)}-1)\varphi_\delta\notag \\
    &\quad+ \big(e^{-\ell(x)}U_te^{\ell(x)}-e^{-\ell_\varepsilon(x)}U_te^{\ell_\varepsilon(x)}\big)(\varphi-\varphi_\delta),\label{eq:strongconv1}
\end{align}
and estimate the $L^2$-norm of each term on the right-hand side separately. Using \eqref{eq:Gronwall-eps}, the third term is bounded by
\begin{align}
   \big\| \big(e^{-\ell(x)}U_te^{\ell(x)}-e^{-\ell_\varepsilon(x)}U_te^{\ell_\varepsilon(x)}\big)(\varphi-\varphi_\delta)\big\|_{L^2}\le C_t\delta, \label{eq:strongconv2}
\end{align}
uniformly in $\varepsilon>0$. To estimate the second term, observing that $-\ell_\varepsilon(x)\le\max(-\ell(x),0)$, we can bound
\begin{align*}
   & \big\|e^{-\ell_\varepsilon(x)}U_te^{\ell(x)}(e^{\ell_\varepsilon(x)}e^{-\ell(x)}-1)\varphi_\delta\big\|_{L^2}\\
   &\le\big\|e^{-\ell(x)}U_te^{\ell(x)}(e^{\ell_\varepsilon(x)}e^{-\ell(x)}-1)\varphi_\delta\big\|_{L^2}+\big\|U_te^{\ell(x)}(e^{\ell_\varepsilon(x)}e^{-\ell(x)}-1)\varphi_\delta\big\|_{L^2}\\
   &\le C\big\|(e^{\ell_\varepsilon(x)}e^{-\ell(x)}-1)\varphi_\delta\big\|_{L^2}+\big\|(e^{\ell_\varepsilon(x)}-e^{\ell(x)})\varphi_\delta\big\|_{L^2},
\end{align*}
for some positive constant $C$, where we used that $e^{-\ell(x)}U_te^{\ell(x)}$ belongs to $\mathcal{B}(L^2)$ and $U_t$ is unitary in the second line. Since $\varphi_\delta$ is compactly supported, we deduce from Lebesgue's dominated convergence Theorem that
\begin{align}
   & \big\|e^{-\ell_\varepsilon(x)}U_te^{\ell(x)}(e^{\ell_\varepsilon(x)}e^{-\ell(x)}-1)\varphi_\delta\big\|_{L^2}\to 0 \quad\text{as}\quad  \varepsilon\to0.\label{eq:strongconv3}
\end{align}
It remains to estimate the first term on the right-hand side of \eqref{eq:strongconv1}. Since (using the bound $-\ell_\varepsilon(x)\le\max(-\ell(x),0)$) we have
\begin{align*}
    &\big|(e^{-\ell(x)}-e^{-\ell_\varepsilon(x)})(U_te^{\ell(x)}\varphi_\delta)(x)\big|\le2e^{-\ell(x)}|U_te^{\ell(x)}\varphi_\delta|(x)+|U_te^{\ell(x)}\varphi_\delta|(x),
\end{align*}
we can apply again Lebesgue's dominated convergence Theorem to deduce that
\begin{align}
    \big\|(e^{-\ell(x)}-e^{-\ell_\varepsilon(x)})U_te^{\ell(x)}\varphi_\delta\big\|_{L^2}\to0 \quad\text{as}\quad \varepsilon\to0.\label{eq:strongconv4}
\end{align}
Putting together \eqref{eq:strongconv1}--\eqref{eq:strongconv4}, we have shown that
\begin{align*}
   \limsup_{\varepsilon\to0} \big\|e^{-\ell(x)}U_te^{\ell(x)}\varphi-e^{-\ell_\varepsilon(x)}U_te^{\ell_\varepsilon(x)}\varphi\big\|_{L^2}\le C\delta.
\end{align*}
Since $\delta>0$ is arbitrary, this concludes the proof.
	\end{proof}

Now we are in position to prove Theorem \ref{th:max-vel}.

\begin{proof}[Proof of Theorem \ref{th:max-vel}]
Using Lemmata \ref{lm:convex} and \ref{lm:bound_domain}, we can write
\begin{equation}
     \|\mathbf{1}_YU_t\mathbf{1}_X\|_{\mathcal{B}(L^2)}
    \le \|\mathbf{1}_Ye^{\ell(x)}\|_{\mathcal{B}(L^2)}\|e^{-\ell(x)}U_te^{\ell(x)}\|_{\mathcal{B}(L^2)}\|e^{-\ell(x)}\mathbf{1}_X\|_{\mathcal{B}(L^2)} \label{eq:MVE1}
\end{equation}
with the following bounds on the terms with the characteristic functions:
\begin{equation}
    \|\mathbf{1}_Ye^{\ell(x)}\|_{\mathcal{B}(L^2)}\le e^{-\frac12\dXY}, \qquad \|e^{-\ell(x)}\mathbf{1}_X\|_{\mathcal{B}(L^2)}\le e^{-\frac12\dXY}.\label{eq:MVE2}
\end{equation}
Thus it remains to estimate the norm of  $e^{-\ell(x)}U_te^{\ell(x)}$. Using Lemmata \ref{lm:unif-bound}, \ref{lm:strongconv} and \ref{lm:bound_domain}, we can pass to the limit $\varepsilon\to0$ in \eqref{eq:Gronwall}, yielding
\begin{align}
    &\big\|e^{-\ell(x)}U_te^{\ell(x)}\varphi\big\|_{L^2}^2=\|\varphi\|_{L^2}^2+2\int_0^t\big\langle e^{-\ell(x)}U_\tau e^{\ell(x)}\varphi \, , \,  \Tzero e^{-\ell(x)}U_\tau e^{\ell(x)}\varphi\big\rangle\mathrm{d}\tau. \label{eq:bound_for_Gronwall}
\end{align}
Since $\|\Tzero\|_{\mathcal{B}(L^2)}\le1$ by \eqref{eq:borne-T0}, we deduce that
\begin{align*}
    &\big\|e^{-\ell(x)}U_te^{\ell(x)}\varphi\big\|_{L^2}^2
    \le\|\varphi\|_{L^2}^2+2\int_0^t\big\| e^{-\ell(x)}U_\tau e^{\ell(x)}\varphi\big\|_{L^2}^2\mathrm{d}\tau.
\end{align*}
Gronwall's Lemma then yields
\begin{align*}
    \big\|e^{-\ell(x)}U_te^{\ell(x)}\varphi\big\|_{L^2}^2\le e^{2t}\|\varphi\|_{L^2}^2.
\end{align*}
Therefore
\begin{equation}
    \big\|e^{-\ell(x)}U_te^{\ell(x)}\big\|_{\mathcal{B}(L^2)}\le e^t.\label{eq:MVE3}
\end{equation}
Putting together \eqref{eq:MVE1}, \eqref{eq:MVE2} and \eqref{eq:MVE3} concludes the proof of the theorem.
\end{proof}

Now we prove Corollary \ref{cor:max-vel-bounds-general-subsets} for non necessarily convex subsets $X$ and $Y$.
\begin{proof}[Proof of Corollary \ref{cor:max-vel-bounds-general-subsets}]Without loss of generality one can assume that $\dXY\geq \sqrt{d}$, by taking $C_d$ such that~$C_d e^{-\sqrt{d}}\geq 1$.
Let $r$ in $(0,\frac{\dXY}{2\sqrt{d}}]$. With $Q_{z}=z+r\big[-\frac{1}{2},\frac{1}{2}\big)^d$ for $z$ in~$\mathbb{R}^{d}$, we set 
\[
Z_{X}=\{z\in(r\mathbb{Z})^{d}\mid Q_{z}\cap X\neq\emptyset\}\,,\quad\textnormal{and}\quad 
Z_{Y}=\{z\in(r\mathbb{Z})^{d}\mid Q_{z}\cap Y\neq\emptyset\}\,.
\]
Then $X\subseteq X_r = \bigcup_{x\in Z_{X}}Q_{x}$ and $Y \subseteq Y_r = \bigcup_{y\in Z_{Y}}Q_{y}$ so
\[
\|\mathbf{1}_{Y}U_{t}\mathbf{1}_{X}\|_{\mathcal{B}(L^{2})} 
= \sup_{\substack{\|f\|_{L^{2}}=1\\ \|g\|_{L^{2}}=1} }|\langle\mathbf{1}_{Y}g,U_{t}\mathbf{1}_{X}f\rangle|
\leq\sup_{\substack{\|f\|_{L^{2}}=1\\ \|g\|_{L^{2}}=1}}|\langle\mathbf{1}_{Y_r}g,U_{t}\mathbf{1}_{X_r}f\rangle|\,.
\]
Using the triangle inequality, the Cauchy-Schwarz inequality, and Theorem~\ref{th:max-vel} we obtain
\[
|\langle  \mathbf{1}_{Y_r}g, U_{t}\mathbf{1}_{X_r}f\rangle| 
  \leq\sum_{\substack{x\in Z_{X}\\y\in Z_{Y}}}\left|\left\langle \mathbf{1}_{Q_{y}}g,U_{t}\mathbf{1}_{Q_{x}}f\right\rangle \right|
  \leq\sum_{\substack{x\in Z_{X}\\y\in Z_{Y}}}\|\mathbf{1}_{Q_{y}}g\|_{L^{2}}e^{t-\mathrm{dist}(Q_{y}\,,Q_{x})}\|\mathbf{1}_{Q_{x}}f\|_{L^{2}}\,.
\]

For~$x$ in~$Z_X$ and~$y$ in~$Z_Y$, the distance between $Q_x$ and $Q_y$ is larger than~$|x-y|-r\sqrt{d}$ and $\dXY-r\sqrt{d}\leq |x-y|$, so we can rewrite the sums so that a convolution product on~$(r\mathbb{Z})^d$ appears, and one can use the Hölder and Young inequalities:
\begin{align*}
|\langle\mathbf{1}_{Y_r}g, U_{t} \mathbf{1}_{X_r}f \rangle |
& \leq e^{t+r\sqrt{d}}\sum_{x,y\in(r\mathbb{Z})^{d}}\|\mathbf{1}_{Q_{y}}g\|_{L^{2}}\,\mathbf{\delta}_{|x-y|\geq \dXY-r\sqrt{d}}\,e^{-|x-y|}\|\mathbf{1}_{Q_{x}}f\|_{L^{2}}\\ \allowdisplaybreaks
 & \leq e^{t+r\sqrt{d}}\,\|\mathbf{1}_{Q_{y}}g\|_{\ell_{y}^{2}((r\mathbb{Z})^{d};L^{2})}\|\mathbf{\delta}_{|z|\geq \dXY-r\sqrt{d}}\,e^{-|z|}\|_{\ell_{z}^{1}((r\mathbb{Z})^{d};\mathbb{R})}\|\mathbf{1}_{Q_{x}}f\|_{\ell_{x}^{2}((r\mathbb{Z})^{d};L^{2})}\\
 & \leq e^{t+r\sqrt{d}}\,\|g\|_{L^{2}}\|\mathbf{\delta}_{|z|\geq \dXY/r-\sqrt{d}}\,e^{-r|z|}\|_{\ell_{z}^{1}(\mathbb{Z}^{d};\mathbb{R})}\|f\|_{L^{2}}\,.
\end{align*}
It remains to estimate $K(X,Y,r)=e^{r\sqrt{d}}\|\mathbf{\delta}_{|z|\geq \dXY/r-\sqrt{d}}\,e^{-r|z|}\|_{\ell_{z}^{1}(\mathbb{Z}^{d};\mathbb{R})}$. For the sake of shortness, the letter~$R$ denotes~$\frac{\dXY}{r}-\sqrt{d}$, and with our assumption on $r$, $R\geq\sqrt{d}$. We have
\[
K(X,Y,r)  = e^{r\sqrt{d}}\sum_{\substack{z\in\mathbb{Z}^{d}\\
|z|\geq R
}
}e^{-r|z|} 
  \leq e^{r\sqrt{d}} \sum_{m=0}^{\infty}\sum_{(m+1)R\leq |z|<(m+2)R}e^{-r(m+1)R}\\
\]
As all the unit cubes with center $z$ in the shell $(m+1)R\leq |z'| < (m+2)R$ are included in the shell $(m+1/2)R\leq |z'|< (m+5/2)R$, it holds
\begin{align*}
K(X,Y,r) 
 & \leq e^{r\sqrt{d}}\sum_{m=0}^{\infty}2R\omega_{d}((m+5/2)R)^{d-1}e^{-r(m+1)R}\\
 & \leq C_{d}e^{r\sqrt{d}} e^{-rR}R^{d}\sum_{m=0}^{\infty}(m+d-1)\cdots(m+1)(e^{-rR})^{m}\,.
\end{align*}
An elementary computation yields, for $|z|<1$,
\[\sum_{m=0}^{\infty}(m+d-1)\cdots(m+1)z^{m}= \frac{(d-1)!}{(1-z)^{d}}\]
and thus
\begin{align*}
K(X,Y,r)
 & \lesssim_{d}e^{r\sqrt{d}}e^{r\sqrt{d}-\dXY}R^{d}\frac{1}{(1-e^{r\sqrt{d}-\dXY})^{d}}\\
 & \lesssim_{d}e^{2r\sqrt{d}}e^{-\dXY}(\frac{\dXY}{r}-\sqrt{d})^{d}\frac{1}{(1-e^{r\sqrt{d}-\dXY})^{d}}\\
 & \lesssim_{d}e^{2r\sqrt{d}}e^{-\dXY}\frac{1}{r^{d}}\frac{(\dXY-r\sqrt{d})^{d}}{(1-e^{r\sqrt{d}-\dXY})^{d}}\\
 & \lesssim_{d}e^{2r\sqrt{d}}e^{-\dXY}\frac{1}{r^{d}}(1+\dXY-r\sqrt{d})^{d}.
\end{align*}
Since we can assume $\dXY\geq \sqrt{d}$, as remarked above, we choose $r=\frac{1}{2}$ and we get:
\begin{align*}
K(X,Y,r)
 & \lesssim_{d}e^{-\dXY} \langle\dXY\rangle^{d}\,,
\end{align*}
which yields the result.
\end{proof}

\section{Unitary propagators}\label{sec:Unitary}

In this section we prove Proposition~\ref{cor:existence-unitary-propagator-for-time-dependent-potential} giving a  sufficient criterion on $(V_t)_{t\in[0,T]}$ for the family $(\jnabla + V_t)_{t\in[0,T]}$ to generate a unitary propagator. We use in particular the following result proved in~\cite[Appendix C]{AmmariBreteaux} (the following proposition is stated in an abstract setting in~\cite{AmmariBreteaux}; to simplify the presentation we only consider a particular  case in the $L^2$ setting, which is sufficient for our purpose).

\begin{prop}[Corollary C.4 in \cite{AmmariBreteaux}]
\label{cor:Am-Bre}
Let $I\subseteq \mathbb{R}$ be a closed interval and let $\{(S_t)_{t\in I},S\}$  be a family of self-adjoint operators on $L^2$ such that:
\begin{itemize}
 \item $S\ge 1$ and for all $t$ in $I$, $S_t\geq 1$,
 \item for all $t$ in $I$, $\mathcal{D}(S_t^{1/2})=\mathcal{D}(S^{1/2})$.
\end{itemize}
Let $(A_t)_{t\in I}$ be a family of symmetric bounded operators in $\mathcal{B}(H^{1/2},H^{-1/2})$  satisfying:
\begin{itemize}
 \item $t\in I\mapsto A_t\in \mathcal{B}(H^{1/2},H^{-1/2})$ is continuous.
\end{itemize}
Assume that there exists a continuous function $f:I\to[0,\infty)$ such that for any $t$ in $I$, we have:\\
(i) for any $\psi\in \mathcal{D}(S_t^{1/2})$,
\begin{equation*}
\big|\partial_t\langle \psi,  S_t \psi\rangle_{L^2}\big|\le f(t) \;\|S_t^{1/2}\psi\|_{L^2}^2\,;
\end{equation*}
(ii) for any $\phi,\psi\in\mathcal{D}(S_t^{3/2})$,
\begin{equation*}
\left|\langle S_t\psi, A_t\phi\rangle_{L^2}-\langle A_t\psi,S_t\phi\rangle_{L^2} \right|\le f(t) \,\|S_t^{1/2}\psi\|_{L^2}\;\|S_t^{1/2}\phi\|_{L^2}.
\end{equation*}
Then the non-autonomous Cauchy problem \eqref{Cauchy-pb-propagateur} admits a unique unitary propagator $U(t,s)$.
Moreover, we have
\begin{equation*}
\|S_t^{1/2}U(t,s)\psi\|_{L^2}\le \mathrm{exp}\Big(2 \;\big|\int_s^t f(\tau) \mathrm{d}\tau\big|\Big) \|S_s^{1/2}\psi\|_{L^2}\,, \quad \forall \, t,s\in I\,.
\end{equation*}
In addition, if there exist $c_1,c_2>0$ such that $c_1 S\leq S(t)\leq c_2 S$ for all $t$ in $I$, then there exists~$c>0$
such that
\begin{equation*}
\|U(t,s)\|_{\mathcal{B}(H^{1/2})}\le c\;\mathrm{exp}\Big(2\big|\int_s^t f(\tau)\mathrm{d}\tau\big|\Big) \;, \quad \forall \,  t,s\in I\,.
\end{equation*}
\end{prop}

We also recall a theorem \cite[Theorem X.17]{reed_sim_vol2}, due to Kato, Lions, Lax, Milgram, and Nelson:

\begin{theo}[KLMN Theorem]
   Let $S$ be a positive self-adjoint operator on a Hilbert space and suppose that $q(\varphi,\psi)$ is a symmetric quadratic form on $\mathcal{Q}(S)$ such that there exist~$a$ in~$[0,1)$, $b$ in $\mathbb{R}$ verifying
   \begin{equation}
       \forall \varphi \in \mathcal{D}(S)\,, \quad |q(\varphi,\varphi)| \leq a \langle \varphi, S \varphi \rangle + b \langle \varphi, \varphi \rangle \,.
   \end{equation}
   Then there exists a unique self-adjoint operator $A$ with $\mathcal{Q}(A)=\mathcal{Q}(S)$ and
   \begin{equation}
       \forall \varphi,\psi \in \mathcal{Q}(A) \,, \quad \langle\varphi, A \psi\rangle = \langle \varphi, S \psi \rangle + q(\varphi,\psi)\,.
   \end{equation}
   The operator $A$ is bounded from below by $-b$ and any domain of self-adjointness for $S$ is a form core for $A$.
\end{theo}

As we use the definition of unitary propagator given in \cite[Appendix C]{AmmariBreteaux} with the Hilbert rigging $H^{1/2}\subset L^2\subset H^{-1/2}$, the KLMN theorem and Proposition \ref{cor:Am-Bre} yield the following theorem for the existence of a unitary propagator.

\begin{prop}\label{theorem:Existence-of-propagator-theoretic-result}
 Let $I\subseteq \mathbb{R}$ be a closed interval and let $(q_t(\varphi,\psi))_{t\in I}$ be a family of symmetric quadratic forms on $H^{1/2}$ such that, for all $t$ in $I$, there exist $a_t$ in $[0,1)$, $b_t$ in~$\mathbb{R}$ with
   \begin{equation}
       \forall \varphi \in H^{\frac12}\,, \quad |q_t(\varphi,\varphi)| \leq a_t   \|\varphi\|_{H^{1/2}}^2 + b_t \|\varphi\|_{L^2}^2  \,.
   \end{equation}
   For each such $q_t$, let $A_t$ be the corresponding self-adjoint operator on $L^2$ obtained through the KLMN theorem with $S=\jnabla$.
   Suppose furthermore that
   \begin{itemize}
       \item $\sup_{t\in I} b_t < \infty$ and
       \item for all $\varphi$ in $H^{1/2}$, $t\mapsto q_t(\varphi,\varphi)$ is differentiable on $I$, with
       \begin{equation}
           \sup_{\substack{t\in I\\ \|\varphi\|_{H^{1/2}}=1}} |\partial_t q_t(\varphi,\varphi)| <\infty\,. \label{eqn:bound-time-derivative-q(t)}
       \end{equation}
   \end{itemize}

Then the Cauchy problem \eqref{Cauchy-pb-propagateur} admits a unique unitary propagator $U(t,s)$.
\end{prop}

\begin{proof}
    To show that the non-autonomous Cauchy problem \eqref{Cauchy-pb-propagateur} has a unique solution, we apply Proposition \ref{cor:Am-Bre} with:
   \begin{itemize}
       \item $S=\jnabla$
       \item $S_t=A_t+C_A$ with $C_A = 1 + \sup_{t\in I} b_t$
       \item $f(t)=\sup\{|\partial_s q_s(\varphi,\varphi)|\mid s\in I\,,\, \|\varphi\|_{H^{1/2}}=1\}$.
   \end{itemize}  
   We first remark that $S=\jnabla\geq 1$ and for all $\varphi\in H^{1/2}$ such that $\|\varphi\|_{L^2}=1$,
   \begin{align*}
       \langle \varphi, S_t\varphi\rangle_{L^2}&=\langle\varphi,A_t\varphi\rangle_{L^2}+C_A\\
       &=\langle\varphi,\jnabla\varphi\rangle_{L^2}+q_t(\varphi,\varphi)+C_A\\
       &\ge(1-a_t)\langle\varphi,\jnabla\varphi\rangle_{L^2}-b_t+C_A\ge1,
   \end{align*}
   since $C_A = 1 + \sup_{t\in I} b_t$ and $a_t<1$ for all $t$ in $I$. Hence $S_t\geq 1$ for all $t$ in $I$, and we have in addition that  $\mathcal{D}(S_t^{1/2})=\mathcal{Q}(S_t)=H^{1/2}=\mathcal{D}(S^{1/2})$. Moreover the assumption on the differentiability of $t\mapsto q_t$ implies that the operators $A_t$ extend to operators $\tilde{A}_t$ in $\mathcal{B}(H^{1/2},H^{-1/2})$ which depend continuously on $t$ in the operator norm topology.
   For all $\varphi\in H^{1/2}$, using  \eqref{eqn:bound-time-derivative-q(t)} we have,
   \begin{equation*}
        \big|\partial_t \langle \varphi,S_t\varphi \rangle_{L^2}\big|=|\partial_tq_t(\varphi,\varphi)| \leq f(t) \leq f(t) \|S_t^{1/2}\varphi\|_{L^2}^2   , 
   \end{equation*}
   since $S_t\ge1$.
Finally, the bound on the commutator is obvious since $A_t$ and $S_t=A_t+C_A$ commute.
\end{proof}

Proposition~\ref{theorem:Existence-of-propagator-theoretic-result} allows us to consider propagators generated by family of operators of the form $(\jnabla + V_t)_{t\in [0,T]}$.

\begin{proof}[Proof of Proposition~\ref{cor:existence-unitary-propagator-for-time-dependent-potential}]Let $I=[0,T]$. 
 The first hypothesis on $V_t$ in the statement of Proposition~\ref{cor:existence-unitary-propagator-for-time-dependent-potential} readily implies that, for all $t\in I$,
    \begin{equation}\label{eqn:condition-kato-reillich}
        \forall\psi\in H^{\frac12}\,, \quad \Big|\int_{\mathbb{R}^d} |\psi|^2  \, V_t \Big| \leq a_t\big\langle \psi, \jnabla\psi\big\rangle+b_t\langle\psi,\psi\rangle,
    \end{equation}
    with $a_t=\|V_{\mathcal{B},t}\|_{\mathcal{B}(H^{1/2},H^{-1/2})}$ and $b_t=\|V_{\infty,t}\|_{L^\infty}$. 
    Using in addition the other hypotheses in the statement of Proposition~\ref{cor:existence-unitary-propagator-for-time-dependent-potential}, it is clear that the quadratic form
    \begin{equation}
        q_t(\varphi,\psi)=\int_{\mathbb{R}^d} \bar\varphi \, V_t \, \psi
    \end{equation}
    satisfies the assumptions of Proposition~\ref{theorem:Existence-of-propagator-theoretic-result}. This shows Proposition~\ref{cor:existence-unitary-propagator-for-time-dependent-potential}.
\end{proof}

\appendix

\section{Sharpness of the maximal velocity estimate for convex subsets}\label{app:sharp}
In this appendix we justify the ``sharpness'' of the maximal velocity estimate proven in Theorem \ref{th:max-vel}, in the sense given in Remark \ref{rk:sharp}. We begin with the following proposition.
\begin{prop}\label{prop:sharp}
Let $0<\delta<1$, $\varepsilon>0$. There exists $C_{\delta,\varepsilon}>0$ such that, for all $t>0$, there exist two convex subsets $X,Y\subset\R^d$ satisfying $\mathrm{dist}(X,Y)=\delta t$ and
\begin{equation*}
    \big\|\mathbf{1}_Y e^{-it\langle\nabla\rangle}\mathbf{1}_X\big\|_{\mathcal{B}(L^2)}\ge 1-\varepsilon-\frac{C_{\delta,\varepsilon}}{t}.
\end{equation*}
\end{prop}

\begin{proof}
   Let $0<\delta<1$. Introducing the notation $\Theta_1:=-i\partial_{x_1}\langle\nabla\rangle^{-1}$, where $x_1$ stands for the first variable in $\mathbb{R}^d$, we note that
   \begin{equation}\label{eq:explicit}
       e^{it\jnabla}x_1e^{-it\jnabla}=x_1+t\Theta_1.
   \end{equation}
   Since the spectrum of $\Theta_1$ is $\sigma(\Theta_1)=[-1,1]$, we can consider $\varphi_\delta\in L^2$, $\|\varphi_\delta\|_{L^2}=1$, such that 
    \begin{equation}\label{eq:def_varphi}
        \mathbf{1}_{(\frac12(1+\delta),1]}(\Theta_1)\varphi_\delta=\varphi_\delta.
    \end{equation}
Now let $\varepsilon>0$ and fix $R_{\delta,\varepsilon}$ such that
    \begin{equation}\label{eq:defR}
        \|\mathbf{1}_{|x_1|\ge R_{\delta,\varepsilon}}\varphi_\delta\|_{L^2}\le\frac\varepsilon2.
    \end{equation}
Let $t>0$ and choose
    \begin{equation*}
        X=X_{\delta,\varepsilon}:=\{x_1\le R_{\delta,\varepsilon}\}, \qquad Y=Y_{\delta,\varepsilon,t}:=\{x_1\ge R_{\delta,\varepsilon}+\delta t\}.
    \end{equation*}
    Let $f$ be a smooth function such that $\mathrm{supp}(f)\subset(0,\infty)$, $0\le f\le 1$ and $f\equiv1$ on $[\frac12(1-\delta),\infty)$. We first write
    \begin{multline}
       \big\|\mathbf{1}_Y e^{-it\langle\nabla\rangle}\mathbf{1}_X\big\|_{\mathcal{B}(L^2)} =\Big\|\mathbf{1}_{[0,\infty)}\Big(\frac{x_1}{t}-\frac{R_{\delta,\varepsilon}}{t}-\delta\Big) e^{-it\langle\nabla\rangle}\mathbf{1}_X\Big\|_{\mathcal{B}(L^2)}\\
        \ge\Big\|f\Big(\frac{x_1}{t}-\frac{R_{\delta,\varepsilon}}{t}-\delta\Big) e^{-it\langle\nabla\rangle}\mathbf{1}_X\Big\|_{\mathcal{B}(L^2)}
        =\Big\|f\Big(\frac{x_1}{t}+\Theta_1-\frac{R_{\delta,\varepsilon}}{t}-\delta\Big) \mathbf{1}_X\Big\|_{\mathcal{B}(L^2)}, \label{eq:last_proof}
    \end{multline}
    where in the last equality we used the unitarity of $e^{-it\langle\nabla\rangle}$ and the explicit formula \eqref{eq:explicit}. 
    Next \eqref{eq:defR} and the fact that $0\le f\le1$ give
    \begin{equation*}
        f\Big(\frac{x_1}{t}+\Theta_1-\frac{R_{\delta,\varepsilon}}{t}-\delta\Big) \mathbf{1}_X\varphi_\delta=f\Big(\frac{x_1}{t}+\Theta_1-\frac{R_{\delta,\varepsilon}}{t}-\delta\Big) \mathbf{1}_{[-R_{\delta,\varepsilon},R_{\delta,\varepsilon}]}(x_1)\varphi_\delta+\mathrm{Rem}_1,
    \end{equation*}
    with $\|\mathrm{Rem}_1\|_{L^2}\le\frac\varepsilon2$. 
    
    Let $F$ be an almost analytic extension of $f$. This means that $F$ belongs to $\mathcal{C}^\infty(\mathbb{C})$, $F|_{\mathbb{R}}=f$, $\mathrm{supp}(F)\subset\{z\in\mathbb{C}\, |\, |\mathrm{Im}(z)|\le C\langle\mathrm{Re}(z)\rangle\}$ for some $C>0$ and, for all~$n$ in~$\mathbb{N}$, $|\frac{\partial F}{\partial \bar z}(z)|\le C_n|\mathrm{Im}(z)|^n\langle\mathrm{Re}(z)\rangle^{-n-1}$, with $C_n>0$. Since $f$ does not decay at $\infty$, we cannot directly use the Helffer-Sj\"ostrand representation. We therefore introduce an artificial cutoff: Let $\eta$ in~$\mathcal{C}_0^\infty(\mathbb{R};[0,1])$ such that $\eta=1$ near $0$ and, for $\Lambda>0$, let $\eta_\Lambda(\cdot)=\eta(\cdot/\Lambda)$. Let~$\tilde\eta$ in~$\mathcal{C}_0^\infty(\mathbb{C})$ be an almost analytic extension of $\eta$ and~$\tilde\eta_\Lambda(\cdot)=\tilde\eta(\cdot/\Lambda)$. Define $f_\Lambda=f\eta_\Lambda$ and~$ F_\Lambda=F\tilde \eta_\Lambda$. In particular $F_\Lambda$ satisfies $|\frac{\partial F_\Lambda}{\partial \bar z}(z)|\le C_n|\mathrm{Im}(z)|^n\langle\mathrm{Re}(z)\rangle^{-n-1}$, uniformly in $\Lambda\ge1$. We can then write (see e.g. \cite{derezinski_ger})
\begin{align*}
	&	f_\Lambda\Big(\frac{x_1}{t}+\Theta_1-\frac{R_{\delta,\varepsilon}}{t}-\delta\Big)-f_\Lambda(\Theta_1-\delta)\\
	&=-\frac1\pi\int \frac{\partial F_\Lambda}{\partial \bar z}(z) \Big(\frac{x_1}{t}+\Theta_1-\frac{R_{\delta,\varepsilon}}{t}-\delta-z\Big)^{-1}\frac{x_1-R_{\delta,\varepsilon}}{t}(\Theta_1-\delta-z)^{-1} \mathrm{d}\mathrm{Re}(z)\mathrm{d}\mathrm{Im}(z) \\
&=-\frac1\pi\int \frac{\partial F_\Lambda}{\partial \bar z}(z) \Big(\frac{x_1}{t}+\Theta_1-\frac{R_{\delta,\varepsilon}}{t}-\delta-z\Big)^{-1}(\Theta_1-\delta-z)^{-1}\frac{x_1-R_{\delta,\varepsilon}}{t} \mathrm{d}\mathrm{Re}(z)\mathrm{d}\mathrm{Im}(z) \\
	&\quad +\frac1{\pi t}\int \frac{\partial F_\Lambda}{\partial \bar z}(z) \Big(\frac{x_1}{t}+\Theta_1-\frac{R_{\delta,\varepsilon}}{t}-\delta-z\Big)^{-1}(\Theta_1-\delta-z)^{-1}[x_1,\Theta_1](\Theta_1-\delta-z)^{-1}\\
	&\qquad\qquad\mathrm{d}\mathrm{Re}(z)\mathrm{d}\mathrm{Im}(z).
\end{align*}
Using the bound on $\frac{\partial F_\Lambda}{\partial\bar z}$ and the fact that the commutator $[x_1,\Theta_1]$ is bounded, we then obtain
    \begin{equation*}
        f_\Lambda\Big(\frac{x_1}{t}+\Theta_1-\frac{R_{\delta,\varepsilon}}{t}-\delta\Big) \mathbf{1}_{[-R_{\delta,\varepsilon},R_{\delta,\varepsilon}]}(x_1)=f_\Lambda(\Theta_1-\delta) \mathbf{1}_{[-R_{\delta,\varepsilon},R_{\delta,\varepsilon}]}(x_1)+\mathrm{Rem}_2,
    \end{equation*}
    with $\|\mathrm{Rem}_2\|_{\mathcal{B}(L^2)}\le \frac{CR_{\delta,\varepsilon}}{t}$, where $C>0$ does not depend on $\delta,\varepsilon,t,\Lambda$. Using \eqref{eq:defR}, we can then rewrite
    \begin{equation*}
        f_\Lambda(\Theta_1-\delta) \mathbf{1}_{[-R_{\delta,\varepsilon},R_{\delta,\varepsilon}]}(x_1)\varphi_\delta=f_\Lambda(\Theta_1-\delta) \varphi_\delta+\mathrm{Rem}_3,
    \end{equation*}
    with $\|\mathrm{Rem}_3\|_{L^2}\le\frac\varepsilon2$, uniformly in $\Lambda\ge1$. Letting $\Lambda\to\infty$ yields
    \begin{equation*}
        \big\|f(\Theta_1-\delta) \mathbf{1}_{[-R_{\delta,\varepsilon},R_{\delta,\varepsilon}]}(x_1)\varphi_\delta-f(\Theta_1-\delta) \varphi_\delta\big\|_{L^2}\le\|\mathrm{Rem}_3\|_{L^2}.
    \end{equation*}
    Finally, since $f\equiv1$ on $[\frac12(1-\delta),\infty)$, we deduce from \eqref{eq:def_varphi} that
    \begin{equation*}
        f(\Theta_1-\delta) \varphi_\delta=\varphi_\delta.
    \end{equation*}
Putting all together, we have shown  that
    \begin{equation*}
        \Big\|f\Big(\frac{x_1}{t}+\Theta_1-\frac{R_{\delta,\varepsilon}}{t}-\delta\Big) \mathbf{1}_X\varphi_\delta\Big\|_{L^2}\ge1-\varepsilon-\frac{CR_{\delta,\varepsilon}}{t}.
    \end{equation*}
    Since $\|\varphi_\delta\|_{L^2}=1$, this concludes the proof of the proposition.
\end{proof}

\begin{cor}\label{cor:sharp}
Let $0<\delta<1$. There exist $t>0$, and $X$ and Y convex subsets
of $\mathbb{R}^{d}$ such that
\[
\|\boldsymbol{1}_{Y}e^{-it\langle\nabla\rangle}\boldsymbol{1}_{X}\|_{\mathcal{B}(L^{2})}>e^{\delta t-\mathrm{dist}(X,Y)}\,.
\]
\end{cor}
\begin{proof}
Let $\tilde{\delta}=(\delta+1)/2$ and $\varepsilon$ in $(0,1)$.
Applying Proposition \ref{prop:sharp} we deduce that for some constant $C_{\tilde{\delta},\varepsilon}>0$,
for all $t>0$ there exist two convex sets $X_{\tilde{\delta},\varepsilon}$
and $Y_{\tilde{\delta},\varepsilon,t}$ such that the equality $\mathrm{dist}(X_{\tilde{\delta},\varepsilon},Y_{\tilde{\delta},\varepsilon,t})=\tilde{\delta}t$ holds, 
and
\[
\|\boldsymbol{1}_{Y_{\tilde{\delta},\varepsilon,t}}e^{-it\langle\nabla\rangle}\boldsymbol{1}_{X_{\tilde{\delta},\varepsilon}}\|_{\mathcal{B}(L^{2})}\geq1-\varepsilon-\frac{C_{\tilde{\delta},\varepsilon}}{t}\,.
\]
Thus for $t$ larger that some $T_{\tilde{\delta},\varepsilon}>0$~,
\[
\|\boldsymbol{1}_{Y_{\tilde{\delta},\varepsilon,t}}e^{-it\langle\nabla\rangle}\boldsymbol{1}_{X_{\tilde{\delta},\varepsilon}}\|_{\mathcal{B}(L^{2})}>e^{\frac{\delta-1}{2}t}=e^{\delta t-\mathrm{dist}(X_{\tilde{\delta},\varepsilon},Y_{\tilde{\delta},\varepsilon,t})}\,,
\]
which is the result.
\end{proof}
\bibliography{biblio}

\doclicenseThis
\end{document}